\documentclass[aip, floatfix, reprint]{revtex4-1}

\usepackage[utf8]{inputenc}
\usepackage[T1]{fontenc}
\usepackage{mathptmx}
\usepackage{amsmath}
\usepackage{amssymb}
\usepackage{amsfonts}
\usepackage{graphicx}
\usepackage{bm}% bold math
\usepackage{multirow}
\usepackage{xcolor}
\usepackage{siunitx}
\usepackage{booktabs}
\usepackage{xr-hyper}
\usepackage{braket}
\usepackage{hyperref}
\hypersetup{
    colorlinks=true,
    citecolor=black,
    linkcolor=black,
    urlcolor=cyan,
}

\begin{document}

\title{Improved decision making with similarity based machine learning: Applications in chemistry}

\author{Dominik Lemm}
\affiliation{University of Vienna, Faculty of Physics,  Kolingasse 14-16, AT-1090 Vienna, Austria}
\affiliation{University of Vienna, Vienna Doctoral School in Physics, Boltzmanngasse 5, AT-1090 Vienna, Austria}
\author{Guido Falk von Rudorff}
\affiliation{University of Vienna, Faculty of Physics,  Kolingasse 14-16, AT-1090 Vienna, Austria}
\author{O. Anatole von Lilienfeld}
\email{anatole.vonlilienfeld@utoronto.ca}
\affiliation{Departments of Chemistry, Materials Science and Engineering, and Physics, University of Toronto, St. George Campus, Toronto, ON, Canada}
\affiliation{Vector Institute for Artificial Intelligence, Toronto, ON, M5S 1M1, Canada}
\affiliation{Machine Learning Group, Technische Universit\"at Berlin and Institute for the Foundations of Learning and Data, 10587 Berlin, Germany}

\date{\today}

\begin{abstract}

Despite the fundamental progress in autonomous molecular and materials discovery, data scarcity throughout chemical compound space still severely hampers the use of modern ready-made machine learning models as they rely heavily on the paradigm, 'the bigger the data the better'. 
Presenting similarity based machine learning (SML), we show an approach to select data and train a model on-the-fly for specific queries, enabling decision making in data scarce scenarios in chemistry.
By solely relying on query and training data proximity to choose training points, only a fraction of data is necessary to converge to competitive performance.
After introducing SML for the harmonic oscillator and the Rosenbrock function, we describe 
applications to scarce data scenarios in chemistry which include 
quantum mechanics based molecular design and organic synthesis planning.
Finally, we derive a relationship between the intrinsic dimensionality and volume of feature space, governing the overall model accuracy.

\end{abstract}

\maketitle
\section{Introduction}

\begin{figure*}[ht]
    \centering
    \includegraphics[width=\textwidth]{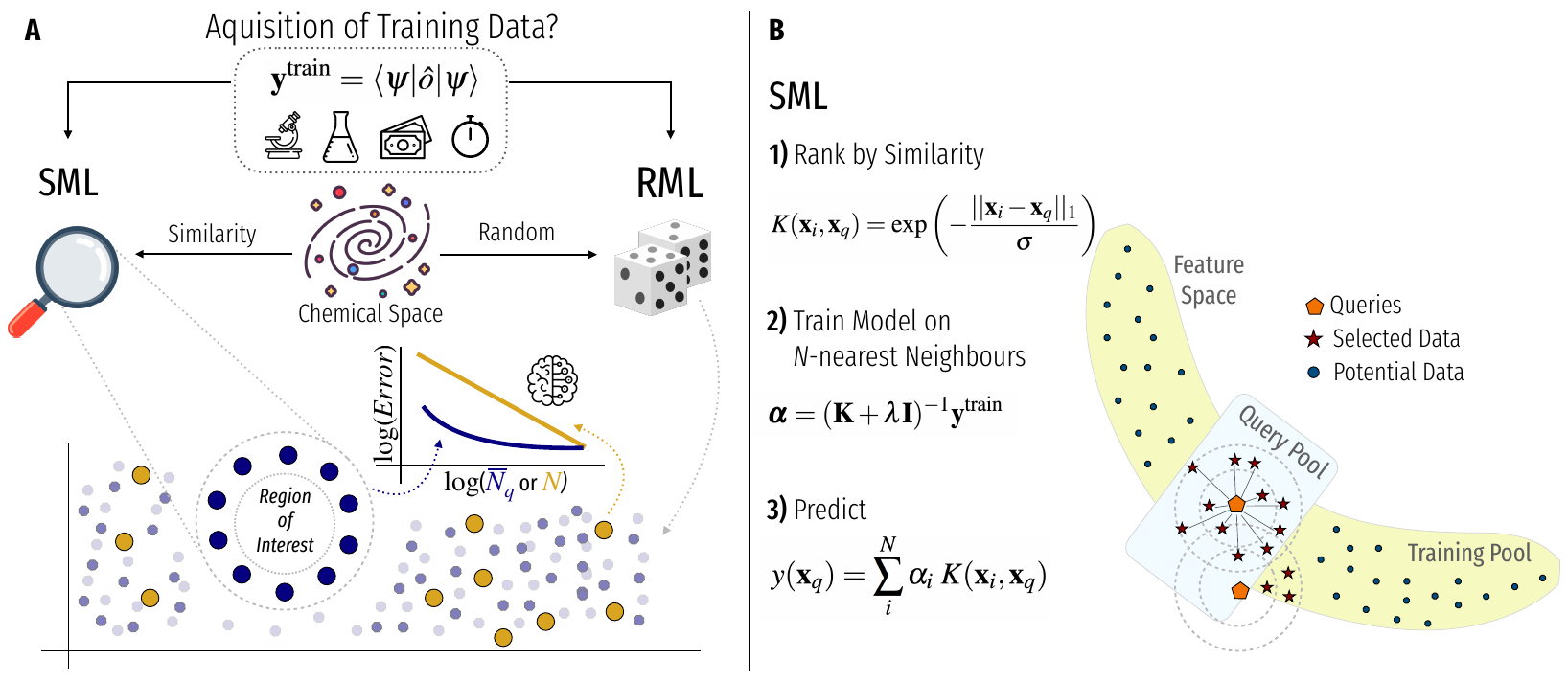}
    \caption{Schematic of similarity based machine learning (SML).
    (\textbf{A}) As a data acquisition strategy, consider similarity based vs.~random selection of costly labels ({\bf y}) corresponding to quantum observables and molecules (represented by  wave-functions {\bf $\Psi$}) from chemical space, resulting in either SML or random machine learning (RML) models, respectively. 
    Due to SML's exclusive focus on the region most relevant to the query improved data-efficiency is achieved, as evinced by learning curves. 
(\textbf{B}) Training procedure of SML with kernel ridge regression: Given a vast pool of potentially interesting queries (rectangle), typically only certain queries are of interest. SML selects the corresponding relevant training points ($\overline{N}_q$) from feature space (banana) as ranked by similarity in representation features (Step No.~1). Consequently, training SML models (Step No.~2) requires costly labels only for the $\overline{N}_q$ nearest neighbours (shown in circles) which do not yet overlap with any other previously selected query. 
Resulting query specific SML models yield similar prediction accuracy for less training data (Step No.~3).}
    \label{fig:banana}
\end{figure*}

Useful answers to experimental design\cite{fisher1936design, chaloner1995bayesian, pukelsheim2006optimal} questions are crucial components for successful decision making\cite{edwards1954theory, pratt1995introduction, berger2013statistical, foster2021statistical, Trommershuser2008} under finite budget and time constraints. 
Conventionally, optimal or near-optimal solutions are proposed by few expert scientists which severely limits humanity’s ‘experimental reach’ as high-dimensional combinatorial solution spaces pose a dramatic bottleneck. Breakthroughs in machine learning and advances in data-availability through high performance computing or high-throughput experimentation are accelerating the pace of scientific discovery and are considered by some to even represent a paradigm-shift\cite{hey2009the, von_lilienfeld_quantum_2018,vonLilienfeld2020}.
In particular, autonomous robotic experimentation in the chemical and biological sciences 
\cite{robotscientist, automationscience, Burger2020, Hse2019,Granda2018, alanbayesian}
revolutionizes humanity's experimental reach through the capacity in which experiments can be executed.
The synergy between machine learning, first principles simulations and experimental design facilitates data driven exploration in molecular and materials discovery through an accelerated feedback loop\cite{EAST}. However, rigorous exploration algorithms are in dire need to bypass the combinatorial wall in the solution space.

With recent interests in data driven machine learning approaches surging, we have set out to investigate how to exploit modern statistical learning in order to assist with such experimental design exploration challenge\cite{NPolitis2017, tye2004application} and decision making problems\cite{edwards1954theory, mldecision}.
A crucial success factor is the availability of accurate machine learning models, often relying on the increasingly prevalent paradigm 'the bigger the data the better' as conveyed through large language model breakthroughs such as OpenAI's impressive Generative Pre-trained Transformer (GPT-)~4.
While GPT-4 is undoubtedly of value through its multimodal capabilities, including novel applications in chemical research\cite{White2023, Jablonka2023, boiko2023emergent} or few shot learning use cases, the overwhelming majority of human endeavours in research are still plagued by intense data scarcity\cite{Weinreich_2023} often with only hundreds if not only dozens of data points being available, without access to pre-trained all purpose models. 

Typically, scarcity is due to high costs imposed by the acquisition of the necessary amount of data. 
Common examples include long simulation times in quantum chemistry\cite{heinen_machine_2020}, cost of compute to perform simulations or money and time to perform laboratory experiments.
Consequently, the development of data-efficient machine learning models and training set selection schemes is crucial\cite{Wen2023, Smith2018, synthtrain, heinim3l, zhang2019active, zubatiuk2021development}.
Within the chemical sciences, a concept frequently referenced in drug discovery research\cite{johnson1990concepts,OBoyle2016,Muratov2020} called `similar structure-similar property principle' (SPP) suggests that when the query is known,
machine learning models should be trained on instances similar to query, rather than on increasingly many diverse ones (Fig.~\ref{fig:banana}A).
Thus, within statistical learning, the training-query similarity emerges as a crucial measure of the predictive power of a model, possibly more relevant than the total quantity of data available.

Here we exploit the concept of SPP in quantum chemical machine learning through introduction of similarity based machine learning (SML) that generates tailored models for arbitrary queries, reaching meaningful prediction accuracy at only a fraction of the total training pool.
Ranking prospective data based on similarity to a query, only the $\overline{N}_q$ nearest neighbours are chosen for training SML models (see Fig.~\ref{fig:banana}B for kernel ridge regression based SML).
Note that if not mentioned otherwise, $\overline{N}_q$ neighbours are selected by proximity rather than all $\overline{N}_q$ within a fixed distance radius. 
While conceptually similar to local learning algorithms\cite{vapnik_local_learning} such as k-nearest-neighbors or moving-least-squares, SML models achieve local neighbourhood weighting through kernel or bagged decision tree based methods.
Within this work, we rely on kernel ridge regression for SML, which in combinations with a Gaussian or Laplacian kernel (Eq.~\ref{eq:gauss}-\ref{eq:laplacian}) can be seen as a form of weighted nearest neighbour regression (see SI for further discussion).
SML would be preferable for applications which require 
a) prior knowledge about the query pool, 
b) few queries of interest with properties that are expensive to measure, e.g. time or money, and 
c) that the property of interest is relatively easy to obtain for similar candidates.

In contrast to other data efficient learning schemes such as active learning, SML focuses on few queries at a time rather than training an all purpose model in an iterative fashion.
Note that the smaller the number of relevant query instances, the more advantageous SML will be. 
However, SML can be extended to iterative active learning by using distance or density based training point selection if a multitude of queries is expected.
Other data efficient learning algorithms such as transfer or few shot learning require pre-trained models, whereas SML solely relies on training points similar to a query, thereby opening the possibility of query aware few shot learning in domains with limited access to pre-trained models.

For exhaustive screening of chemical compound space, by contrast, conventionally
trained random machine learning (RML) models are more appropriate. 
Note that within this work SML and RML are both
based on kernel ridge regression, however, differ in their training data selection and applicability. 
As such, consider multi-objective 
compound design in which an RML model
is used as a filter to scan vast domains of chemical space\cite{kirkpatrick_chemical_2004}, thereby requiring a large and diverse training set in order to answer any query \cite{GmezBombarelli2016, Westermayr2023}. 
Once a candidate compound has been identified,
i.e. we are aware of the query, SML can subsequently be used
to provide highly accurate estimates for properties not covered
by the highly general RML model.

After introducing SML in this article, we demonstrate its usefulness for two distinct applications:
i) Quantum mechanics based molecular design:
Statistical learning of quantum properties holds the promise to accelerate the computational materials design process via fast and accurate machine learning models\cite{gdml,bartok2017}.
Due to the size of chemical compound space (encompassing at least 10$^{60}$ compounds\cite{kirkpatrick_chemical_2004}) and the computational cost of atomistic simulations (ranging from hours to months\cite{heinen_machine_2020}), 
SML can dramatically reduce training data needs if target query compounds are known. 
ii) Organic synthesis planning:
The bottleneck of molecular and materials discovery is amplified in experimental settings, due to large costs and the required manual labour. Additionally, compounds have to be synthesised or purchased before properties can be measured, adding to the cost-time complexity of molecular and materials design. We showcase how SML can be used to efficiently predict relevant properties for target boutique compounds without having make the expenditure necessary to acquire them. 

Finally, we analyse the link between query and training data proximity, allowing to derive a relationship between the intrinsic dimensionality and volume of the feature space spanned by the training data.
The results further corroborate the idea that with increasing local nearest neighbour density an decreasing fraction of the total data pool is required to converge to competitive performance.

\section{Results}

\subsection{Similarity Machine Learning}
\begin{figure}[ht]
    \centering
    \includegraphics[width=0.5\textwidth]{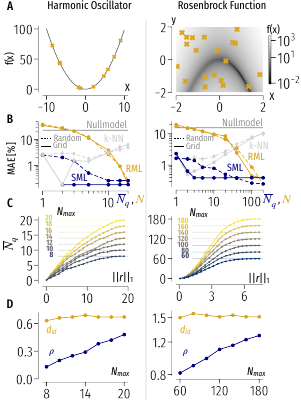}
    \caption{
    Left column:
    SML of a harmonic potential. 
    Right column: SML of the Rosenbrock function. 
    (\textbf{A}) Illustration of target functions and some random query points. 
    (\textbf{B}) Learning curves showing systematic improvement with increasing training set size using RML and SML compared to k-NN baselines.
    Random (dashed) and grid (solid) indicate the initial training pool sampling of the function, respectively.
    The nullmodel corresponds to the averaged function value in training set.
    (\textbf{C}) Average number of nearest neighbours $\overline{N}_q$ as a function of a radius around a query for increasing total training set size $N_{max}$, drawn from respective ranges at random. 
    (\textbf{D}) Intrinsic dimensionality $d_{id}$ and density of points $\rho$ for increasing $N_{max}$ (see Eq.~(\ref{eq:dimfit})).}
    \label{fig:toy}
\end{figure}

To assess the effect of the SML approach on learning, we compare learning curves for random and similarity based machine learning models applied to the analytically known harmonic oscillator and the Rosenbrock functions (Fig.~\ref{fig:toy}~A and B).
The SML learning curves exhibit dramatically lower offsets and faster convergence when compared to randomly selected ML (RML) models.
See methods for more technical details on training and testing.

On average, only very few yet similar training points are necessary for SML to be competitive with the RML model trained on an order of magnitude more training points. 
In comparison with k-nearest neighbour (k-NN) regression, SML shows on a par performance when the training pool was grid sampled (Fig.~\ref{fig:toy}~B).
However, SML outperforms k-NN in randomly sampled training pool settings, suggesting robust performance even in sparsely sampled regions.
As mentioned before, since SML is trained on the fly for each query, the use of SML is only advantageous if few queries are of concern. For exhaustive screening and enumeration studies, the conventional RML approach is more appropriate. 
We note the difference in shape of an SML curve when compared to learning curves reported in the literature.
The rapid onset of convergence indicates that only a fraction of available nearest neighbours are necessary to reach the maximum predictive accuracy.
Note that in the limit of maximum dataset size, the SML model's prediction error will always converge towards the corresponding RML prediction error.
\par

\subsection{Dimensionality analysis}
We have analysed the SML models throughout this paper using learning curves.
The relationship between the predictive accuracy of a model and the number of training points is a fundamental concept in statistical learning. If trained properly, the learning curve\cite{cortes_learning_1994} must reflect the inverse relationship between out-of-sample prediction (test) error as a function of number of training points $N$. 
Note, in the case of SML based learning curves, $N$ corresponds to the number of nearest-neighbours $\overline{N}_q$ chosen for training.
 The logarithmized version of the leading error term is,
\begin{equation}
  \log(\textrm{Error}) \approx \log(a) - b~\log(N),
    \label{eq:lrc2}
\end{equation}
with the slope $b$ being related to the intrinsic dimensionality ($d_{id}$) and $a$ being related to the target similarity\cite{learningcurve_review}.
For the development of data efficient machine learning models it is desirable to understand what decreases $\log(a)$ and increases $b$.
Efforts in estimating and analysing the impact of $d_{id}$ have shown an inverse correlation with high model performance, implying that low  $d_{id}$ data is generally easier to learn\cite{dimlearning, NEURIPS2019_cfcce062, DBLP:conf/iclr/PopeZAGG21},
and that compact representations suffice.
In the limit of large $N$ and randomly sampled data, Eq.~\ref{eq:lrc2} must become linear for Gaussian Process Regression\cite{cortes_learning_1994} as well as neural network regressors\cite{Mller1996}.

\begin{figure*}[ht]
    \centering
    \includegraphics[width=\textwidth]{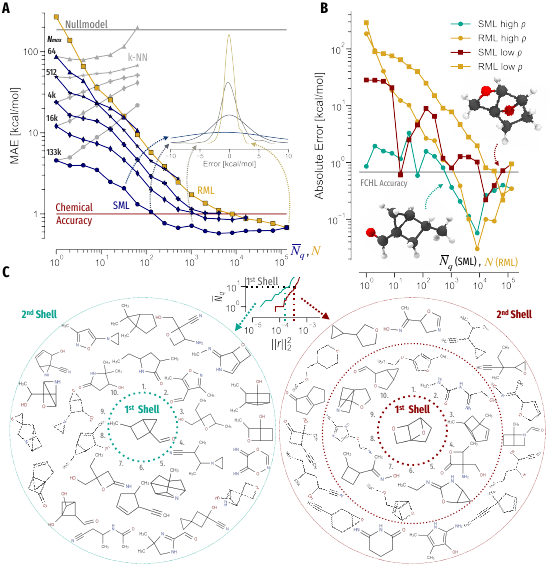}
    \caption{Training and analysis of SML models of atomization energies of QM9~\cite{ramakrishnan_quantum_2014} compounds using the FCHL19~\cite{christensen_fchl_2020} representation. 
    (\textbf{A}) Learning curves for RML, SML and k-NN for increasing maximal training set sizes, drawn at random. 
    The inset depicts the error distribution of SML models at varying training set sizes for the largest $N_{max}$. 
    Chemical accuracy threshold shown as horizontal line.
    The nullmodel corresponds to the averaged function value in training set.
    (\textbf{B}) Learning curves of RML and SML models for predicting 4-methylbicyclo[2.1.0]pentane-2-carbaldehyde (\url{https://tinyurl.com/2p9dtueh}) and 3,8-dioxatricyclo[4.2.1.0$^{2,5}$]nonane (\url{https://tinyurl.com/2p8e4d3h}) with either high or low nearest neighbour density, respectively. 
    (\textbf{C}) Display of the first shell (closest ten molecules) and second shell (next closest fifteen) neighbours for the two query compounds.
    }
    \label{fig:qm9}
\end{figure*}

\begin{figure*}[ht]
    \centering
    \includegraphics[width=\textwidth]{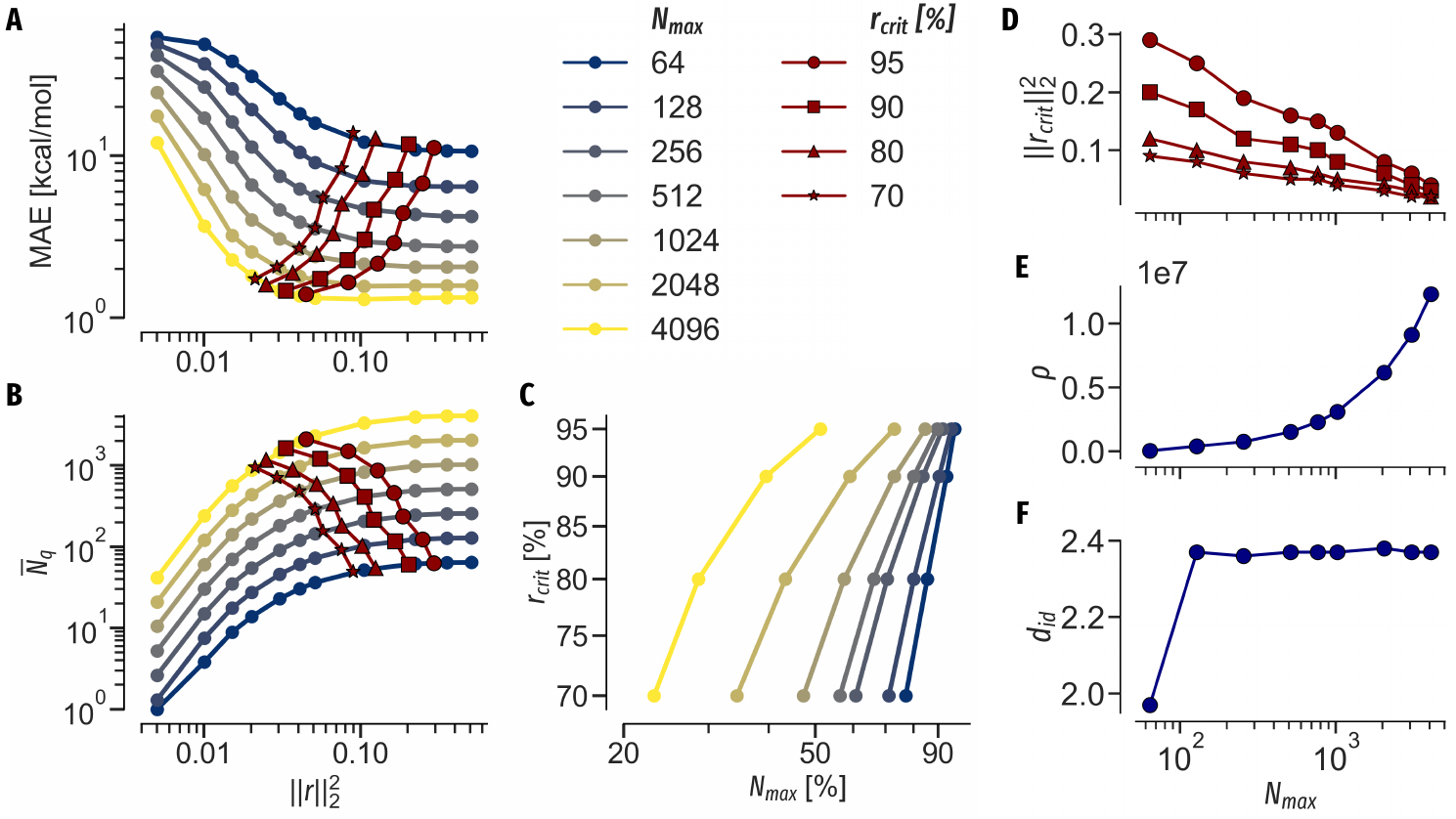}
    \caption{Distance based analysis of SML models of learning atomization energies of QM9 using FCHL19 and a Gaussian kernel. 
    $r_{crit}$ represents the distance radius at which a certain percentage of the maximally attainable predictive accuracy is reached.
    (\textbf{A}) Systematic learning of atomization energies using only training instances within the L2-Distance specified on abscissa.  
    (\textbf{B}) Averaged number of nearest neighbours around a query at different total $N_{max}$ for increasing L2-distance radius. 
    (\textbf{C}) Percentage of data set size necessary to reach a certain accuracy as determined via $r_{crit}$.
    (\textbf{D-F}) $r_{crit}$, density and intrinsic dimensionality ($d_{id}$) of the QM9\cite{ramakrishnan_quantum_2014} chemical space as a function of available training points, respectively. 
    }
    \label{fig:qm9_dim}
\end{figure*}

\begin{figure}[hp]
    \centering
    \includegraphics[width=\linewidth]{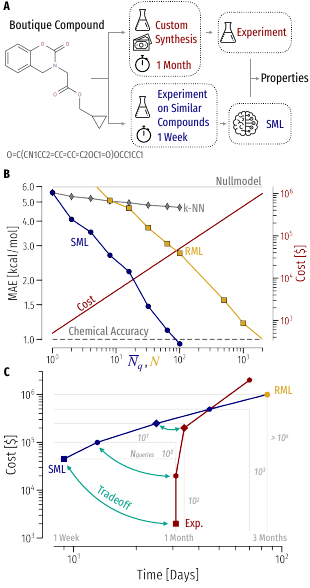}
    \caption{Example of how to use SML within a virtual custom synthesis scenario.
    (\textbf{A}) Suppose, the decision to make an investment in a certain desired boutique query compound (exemplary molecular graph and string shown as inset \url{https://tinyurl.com/5yvxxzst}) depends on a certain property. 
    Rather than  ordering a costly and time-intensive custom synthesis, one can instead simply predict the property by first ordering sufficient more easily available compounds which are similar to the query, and by subsequently measuring their properties in order to train an accurate SML model.  
    The nullmodel corresponds to the averaged function value in training set.
    (\textbf{B}) Respective learning curves and training data acquisition cost for RML and SML models of an exemplary property (free energies of solvation) using molecules drawn at random from Enamine REAL libraries~\cite{enamine_real,enamine_realdb}. Dashed horizontal line indicates chemical accuracy threshold (1~kcal/mol corresponding to experimental uncertainties in thermochemical combustion).
    (\textbf{C}) Cost vs. time associated with the generation of an SML or RML dataset ($\sim$500 US\$/compound) in comparison to a ball-park synthesis cost of 2'000 US\$/query for different number of queries  $N_{queries}$ to be expected. 
    The time estimate includes approximate delivery times of 5 days for SML/RML compounds and 1 month for boutique compounds, respectively, as well as an experimental throughput of 25 compounds per day.
    }
    \label{fig:enamine}
\end{figure}

Calculating the distances of all training points in a pool $N_{max}$ to all queries, respectively, the average amount of nearest neighbours $\overline{N}_q$ within a certain distance radius $r$ can be determined.
Fig.~\ref{fig:toy}C confirms the intuition that by increasing the total data pool $N_{max}$, the local density, i.e. the amount of neighbours within a fixed radius, grows as well.
To quantify the relationship between $N_{max}$ and distance radius $r$, we use the definition of density $\rho$, i.e. 
$    \rho = N/V  = N/{r^{d_{id}}}
$,
with volume $V$, distance radius $r$ and intrinsic dimensionality $d_{id}$.
To obtain values for $\rho$ and $d_{id}$, we transform the density equation to fit the steepest slope of $\overline{N}_q$ as a function of distance $r$ on a double logarithmic scale (Fig.~\ref{fig:toy}C), yielding:
\begin{equation}
        \log(N) = \log(\rho) + d_{id}~\log(r),
    \label{eq:dimfit}
\end{equation}

Fig.~\ref{fig:toy}D shows the resulting density $\rho$ to increase linearly with larger pool-sizes $N_{max}$.
Interestingly, our calculated $d_{id}$ effectively underestimates the respective formal dimensionality of 1D and 2D for the harmonic oscillator and Rosenbrock function (Fig.~\ref{fig:toy}D). 
Our analysis indicates that edge artifacts introduced through the fixed function boundaries, as depicted in Fig.~\ref{fig:toy}A, contribute to this effect:
Queries that are close to the boundary will have less neighbours when increasing the distance radius than queries in the center, leading to an underestimated  $d_{id}$ (see SI, Fig.~S1). Given an infinite space and perfectly centered points, the formal $d_{id}$ can be recovered.
We note that other nearest neighbour or maximum likelihood  $d_{id}$ estimations\cite{PhysRevLett.130.067401, Majumdar2016, amsaleg2015estimating, pettis1979intrinsic} are possible, but can suffer from negative bias through edge effects in higher dimensions\cite{Facco2017, mle_id}. \par
Having introduced the concept of SML and learning curves on mathematical toy functions, we will now apply SML to learning quantum properties of chemical compounds.

\subsection{Quantum mechanics based virtual compound design}

To illustrate a meaningful use-case of SML, consider a typical problem of computational molecular and materials design: Organic light-emitting diodes.
Consider furthermore that a promising compound would have been identified through sophisticated screening of a very large set of potential candidate compounds.
Suppose that the screening imposes all necessary constraints on the candidate compound, except for quantum mechanics based excited states calculations which is not only difficult to model, but is also computationally expensive for larger compounds\cite{Westermayr2020, Westermayr2021}.
In this scenario, it is difficult to make a rational decision due to the high computational cost and time required to perform the quantum calculation. Furthermore, given the scarcity of literature excited states data throughout chemical compound space, RML models such as state of the art neural network architectures\cite{pmlr-v139-satorras21a, Atz2022, EquiformerV2, 2022equivariant} in quantum chemistry would potentially require larger volumes of data to yield accurate estimates.
By contrast, due to its data-efficiency, SML will be able to produce substantially more accurate predictions after training on substantially less data.

Obviously, this molecular design scenario is not restricted to organic electronics. As a numerical illustration, we report the application of SML to learning quantum mechanics based calculated atomization energies (rather than excited states) of organic molecules, as reported in the QM9\cite{ramakrishnan_quantum_2014} dataset, rather than for organic materials candidates. 
As for the harmonic oscillator and Rosenbrock function discussed above, Fig.~\ref{fig:qm9}A displays overwhelming numerical evidence that SML models also exhibit superior data-efficiency when it comes to molecules and their properties. 
More specifically, for all tested $N_{max}$, SML learning curves exhibit a much reduced offset, and converge to the accuracy of the RML learning curve at $N_{max}$ for training set sizes that are smaller by orders of magnitude.
In contrast, k-NN did not improve beyond the first neighbour.
Note also the robustness of SML: As the number of nearest neighbours $\overline{N}_q$ grows, its predictive variance decreases systematically (Fig.~\ref{fig:qm9}A inset).
Furthermore, SML performs consistently when applied to quantum properties in chemical compound space as datasets and molecular representations are being varied (for further examples see SI and Fig.~S3).

At increasingly large $N_{max}$, a shoulder becomes apparent for small training set sizes with less than $\overline{N}_q$=30. 
Moreover, the larger $N_{max}$, the more data-efficient the SML model (earlier flattening) suggesting that the increase in local neighbour density leads to faster saturation in learning (see II.~B for in-depth dimensionality analysis in analogy to the harmonic oscillator and Rosenbrock function).
At the largest $N_{max}$ possible within the QM9 data-set ($\sim$134k molecules for training), SML reaches the coveted chemical accuracy of on average 1~kcal/mol at only 100 training points for any given query.
For RML models, this would have solved the 'QM9-IPAM-Challenge' in quantum chemical machine learning\cite{vonLilienfeldburke2020, von_lilienfeld_exploring_2020}.
Note that many applications in quantum mechanics based simulation of molecules and materials are rather concerned with few queries than many.
This indicates, that SML might be the superior choice, than learning a general purpose property throughout chemical compound space, such as a universal force field. 

To illustrate the effect of the query-specific density on the SML model, we compare two query compounds with respectively low and high density of neighbours in chemical space. 
More specifically, we consider the caged ether-bridged C$_6$H$_8$O$_2$ compound,
3,8-dioxatricyclo[4.2.1.0$^{2,5}$]nonane\cite{STOUT, leruli},
and the fused C$_7$H$_{10}$O aldehyde,
4-methylbicyclo[2.1.0]pentane-2-carbaldehyde\cite{STOUT, leruli},
encoding motifs that are respectively rather rare or frequent within the chemical space spanned by QM9\cite{ramakrishnan_quantum_2014}.
SML learning curves (Fig.~\ref{fig:qm9}B)
confirm the expectation: The higher the density of neighbours
the better the learning curve's off-set and slope in the low-data limit. 
Note that for query \textit{specific} training resulting learning curves can also increase with training set size. This is in stark contrast to the statistical learning tenet that \textit{averaged} prediction errors must decay systematically~\cite{cortes_learning_1994,Mller1996}.
Visual inspection of the closest nearest neighbours in terms of FCHL based distances (Fig.~\ref{fig:qm9}C) shows barely any relative chemical trends of similarity. Using BoB based distances, however, 
the visual inspection is more straightforward: Fig.~S5 reveals for example that for a high-density query a tautomer in QM9 got selected by SML, resulting in near-perfect predictions.

\subsection{Dimensionality analysis of QM9}
To gain a better understanding of the relation between query and training data proximity, SML models have been trained~-~similar as reported for the harmonic oscillator and Rosenbrock function - using an increasing distance radius as a measure to select training data.
Fig.~\ref{fig:qm9_dim}A indicates that with increasing $N_{max}$ the distance radius required to reach the final accuracy becomes smaller. 
Concurrently, the average amount of nearest neighbours $\overline{N}_q$ in a smaller radius also increases (Fig.~\ref{fig:qm9_dim}B). 
We define a critical radius $r_{crit}$ to determine the distance cutoff at which 70, 80, 90, 95\% of the maximum accuracy is reached at each $N_{max}$ (Fig.~\ref{fig:qm9_dim}A), respectively. 
We observe that at increasing $N_{max}$, $r_{crit}$ becomes smaller (Fig.~\ref{fig:qm9_dim}D).
Calculating the average amount of neighbours $\overline{N}_q$ within $r_{crit}$ (Fig.~\ref{fig:qm9_dim}B), a pareto front becomes apparent, confirming that at increasing $N_{max}$ a fraction of the data is sufficient (Fig.~\ref{fig:qm9_dim}C).
We apply Eq.~\ref{eq:dimfit} on Fig.\ref{fig:qm9_dim}B to obtain the density $\rho$ and $d_{id}$.
As expected, the density $\rho$ increases with $N_{max}$ (Fig.~\ref{fig:qm9_dim}E).
However, in consideration of the size of chemical compound space and ways of systematic enumeration, it is unclear how and if the density will converge and how this will influence the learning of properties in chemical compound space.
The estimated $d_{id}$ converges to a value of 2.4 for 
$N_{max}>$ $\sim$100 using a Gaussian kernel (Fig.~\ref{fig:qm9_dim}F) and 1.7 using a Laplacian kernel (see SI Fig.~S11). 
While estimates of the $d_{id}$ for popular chemical datasets have not been rigorously reported yet, we note that the $d_{id}$ of a fixed segment of chemical compound space is bound to an upper limit of 4$N$-6 when using a roto-translational invariant representation. 
Reasons for potential underestimates can be finite boundaries in terms of representation or even chemical space. 
Points in high dimensions tend to be localized at the surface of the hypersphere. 
The effect of edge artifacts can further be amplified through the high dimensionality of molecular representation, which typically are vectors containing thousands of entries. 
However, we report that even a high dimensional representation such as BoB ($\sim$1'000 entries for QM9) shows high co-linearity compared to uniform random vectors of the same size (see SI Fig.~S12).

\subsection{Organic synthesis planning}
A SML use-case in the context of managing synthesis planning\cite{Molga2021, Coley2018, Levin2022, MikulakKlucznik2020} within industrial chemistry could be as follows:
Consider the problem of acquiring non-trivial molecular properties of one boutique compound through experimental measurements. 
Synthesis of the compound prior to measurements is also assumed to first require substantial experimental efforts resulting in very long delivery times (also known as 'lead times')\cite{Hughes2011}. 
Note that this is not an unusual 'academic' situation as the latter can nowadays be performed  commercially through  synthesis service companies such as {\tt Enamine Ltd}. 
While such services have reached considerable reliability, reporting success rates of 60-80\% and higher, challenging boutique target compounds, for example cyclopropylmethyl2-(2-oxo-3,4-dihydro-2H-1,3-benzoxazin-3-yl)acetate\cite{STOUT, leruli}
on display in Fig.~\ref{fig:enamine}A can impose substantial time delays until the compound has been made, characterized, and shipped, increasing lead times to several weeks or more. 
Alternatively, SML can be used to make accurate property predictions, rather than measurements, \textit{before} having to wait out the lead time and measuring the target compound's properties: 
First ordering and measuring the most similar compounds with much reduced lead times,  for example readily deliverable (3-5 days) building blocks from the Enamine REAL library\cite{enamine_real, enamine_realdb} used for the synthesis of our exemplary boutique compound,
would allow for rapidly making the measurements to generate the training data for subsequent SML model generation. 
The resulting SML models can then be used to accurately infer the property for the original boutique target compound. 
In comparison to the naive approach of simply waiting for the compound to be delivered, followed by measurements of the properties, this procedure would effectively enable a dramatic cut down of the time to decision while keeping the financial burden for data acquisition at a minimum.

To numerically illustrate the application of SML towards such use cases, we have relied on the aforementioned specific example product (cyclopropylmethyl2-(2-oxo-3,4-dihydro-2H-1,3-benzoxazin-3-yl)acetate),  as well as 999 other boutique molecules (see data availability section for a complete list) with lead times of 3 to 4 weeks on average. 
As a relevant property, we have selected calculated free energy of solvation estimates\cite{RMG, RMG2} as a proxy to the measurement. 
As to be expected from our discussions in the preceding sections, resulting learning curves suggest that on average SML requires a magnitude less data than RML and, similarly, outperforms k-NN (Fig.~\ref{fig:enamine}B).
Assuming  a 500~US\$/compound price level on average, estimated purchasing costs for training data-acquisition are shown, also indicating substantial potential for savings, i.e.~time \textit{is} money. 

Given that there are multiple queries (up to 1'000), we have also assessed to which degree SML can benefit from synergies due to overlapping nearest neighbours of distinct queries. 
This possibility would reduce the scaling from the worst case upper bound imposed by RML, rendering SML  preferable in terms of time and cost for larger numbers of target queries.
To illustrate this scenario, and without loss of generality, we define the target accuracy to correspond to chemical accuracy (1~kcal/mol).
Assuming up to 1'000 queries, we have applied SML to select a jointly combined training set containing the closest nearest neighbours of the selected queries.
Analysis of the $N_{neighs}$ required per query to reach 1~kcal/mol (see SI), indicates that less $N_{neighs}$ are required when considering shared neighbourhood SML models, rather than single query SML.
Note that with increasing $N_{queries}$, also increasingly larger SML training set is required, ultimately converging to the upper bound used within RML models. 
Thus, variety in the query-set also requires variety training set, whereas few queries are more efficiently predicted via SML.

Additionally, time and cost are constraints to be considered during the design process (Fig.~\ref{fig:enamine}C).
Albeit the data-acquisition costs of machine learning approaches can be more costly than directly ordering the target compounds (we assume 2'000~US\$/target compound price in Fig.\ref{fig:enamine}C), SML and RML can become financially advantageous for large sets of $N_{queries}$. 
For a modest number of expected queries, SML will be preferable, whereas for many queries RML is the method of choice.
In our example, however, the cost per query has not been included as an additional constraint when selecting nearest neighbours. This could lower the total cost even further. In terms of time to decision, SML will always be preferable. 
Note that the specific variables cost and time strongly depend on each scenario. 
In conclusion, given additional constraints and a large chemical space to choose from, approximate nearest neighbour selection\cite{ann} is beneficial for reducing cost and time to decision in planning the management of chemical synthesis.

\subsection{Limitations}

We want to note that any limitations that apply to an RML model will also apply to SML since both only differ in their training data, but not in their underlying machine learning algorithm, e.g. kernel ridge regression. 
Since the choice of representation and distance metric influences the choice of nearest neighbours, an improvement in both, e.g. by including more physical knowledge or metric learning\cite{Fabregat2022}, will result in lower learning curve offsets (see SI Fig.~S3).
Conversely, low local nearest neighbour density (Fig.\ref{fig:qm9}B) will require more data in order to reach optimal accuracy. 
Additionally, the compute time for an exact nearest neighbour search grows linearly with $O(n)$. 
However, since the local nearest neighbour density is proportional to the size of search space (see Fig.~\ref{fig:qm9_dim}), the $O(\log{}n)$ scaling of approximate nearest neighbour algorithms, commonly used with millions of data points, becomes more effective. 
Further strategies could encompass to use efficient low dimensional representations for approximate nearest neighbour searches while using higher quality representations for regression, resulting in a time vs. accuracy trade-off.
To make the most out of SML within multi query scenarios, it is recommended to maximize the overlap of shared SML training sets as described in the supplementary information.
Methods such as Auto3D\cite{auto3d}, OQML\cite{heini_geom}, Graph-To-Structure\cite{Lemm2021} or generative models\cite{pmlr-v162-hoogeboom22a, pmlr-v202-xu23n, jing2022torsional} could further facilitate the generation of 3D input structures
for quantum chemistry based models to avoid performance loss by using lower levels of theory geometries such as from force fields\cite{VazquezSalazar2021}.
Additionally, the atoms-in-molecule\cite{AMONS} (AMONS) approach could further aid the applicability of SML by using smaller molecular building blocks to predict significantly larger query compounds.

\section{Discussion}

We have presented numerical evidence that similarity based machine learning (SML), 
an approach to select data and train a model on-the-fly for specific queries, can offer substantial improvements in data efficiency for certain use-cases.
Applying SML to problems in quantum mechanics and organic synthesis planning, we have found dramatically lower offsets in learning curves\cite{cortes_learning_1994}, enabling meaningful decision making at a fraction of the total training set size.
Comparisons of SML with other similarity based methods indicate robust performance, even in high dimensional feature spaces.
Large training set sizes and associated costs of data acquisition render the use of one-fits-all ML models unfeasible related to the size of chemical compound space.
The shape of SML learning curves suggest meaningful deviations from 'the bigger the better' paradigm indicating that learning is rather dominated by locality effects, and even saturates upon inclusion of sufficiently many nearest neighbours in training.

After introducing and demonstrating SML for harmonic oscillator and Rosenbrock functions, we have demonstrated its usefulness for quantum mechanics based molecular design where we can reach the coveted 1~kcal/mol chemical accuracy already after just 100 nearest neighbours. This kind of task requires conventional RML models to be trained on thousands of data points.  
More fundamentally, however, we have derived a relationship between the intrinsic dimensionality and volume of the feature space spanned by the training data which governs the overall model accuracy. 
In managing the planning of experimental organic synthesis projects, we have shown that SML can bypass time intensive custom synthesis by using readily available compounds similar to the boutique compounds of interest. 

We believe that for certain decision problems in chemistry the paradigm of 'the bigger the data the better' is not optimal, and that tailored, query specific SML models reach predictive power already for training set sizes that are orders of magnitude smaller than for conventional RML models. 
More generally, we expect SML to be advantageous for any problem with
a) prior knowledge of and access to molecular features in query pool (to easily calculate similarity), 
b) few queries of interest with labels that are expensive to measure (e.g. long computation times or high costs)
c) labels for similar candidates in pool that are less difficult to obtain.
Moreover, the implicitly defined domain of applicability of the model provides an indication for the reliability of a machine learning model for unseen queries and inherently limits the usefulness outside of this domain (extrapolation\cite{extrapol_ml}).
Potential applications beyond chemistry could include predictive maintenance tasks in computer aided infrastructure\cite{materialsdegrad} or aerospace engineering\cite{aviation1, aviation2}.
Future work will address, automated experimental design, deriving ready made models in quantum chemistry or multilevel learning, improving similarity measurements, as well as exploring fast and approximate neighbour searches (cf.~turbo similarity fusion)\cite{Fabregat2022, Gardiner2009, MirandaQuintana2021-1, MirandaQuintana2021-2}.

\section{Methods}

\subsection{Kernel Ridge Regression (KRR)}

We rely on kernel ridge regression (KRR)\cite{krige_statistical_1951} due to its successes in small data scenarios, however, other machine learning regressors are possible. 
Kernel based methods belong to the class of supervised machine learning models and 
are -- despite their simplicity -- a powerful approach for learning molecules and materials properties.
In ridge regression, a mapping function is learned, relating a representation vector $\textbf{x}$ to a label $y$\cite{krige_statistical_1951, vapnik_nature_2000}.
The 'kernel trick' renders the problem linear through the use of kernel functions that yield the inner product of high dimensional representations $\textbf{x}$.
In practice, KRR is formulated as follows:

\begin{equation}
    y(\textbf{x}_{q}) = \sum^N_i \alpha_i \;  K(\textbf{x}_i, \textbf{x}_{q})
    \label{eq:krr}
\end{equation}

with $\alpha_i$ being the regression coefficient, $\textbf{x}_i$ being the i-th representation vector of the training set, $\textbf{x}_q$ being the query representation and kernel function $K$. 
The regression coefficients $\alpha$ are calculated through a closed-form solution:

\begin{equation}
   \bm{\alpha} = (\textbf{K} + \lambda \textbf{I})^{-1} \textbf{y}^{\textrm{train}}
   \label{eq:train}
\end{equation}

with the identity matrix \textbf{I} and a regularization coefficient $\lambda$. 
The latter depends on the anticipated noise in the input and labels and has to be determined through hyperparameter optimization.

\subsection{Kernel Functions}
As a similarity measure, the Laplacian (Eq.~\ref{eq:laplacian}) and Gaussian (Eq.~\ref{eq:gauss}) kernels are the most common choices. 

\begin{equation}
    K(\textbf{x}_i, \textbf{x}_q) = \exp\left(-\frac{|| \textbf{x}_i - \textbf{x}_q||_1}{\sigma}\right)
    \label{eq:laplacian}
\end{equation}

\begin{equation}
     K(\textbf{x}_i, \textbf{x}_q) = \exp\left(-\frac{||\textbf{x}_i - \textbf{x}_q||^2_2}{2\sigma^2}\right)
    \label{eq:gauss}
\end{equation}

with $\sigma$ being the kernel width, an additional hyperparameter to optimize.
The kernel function for local atomic representations can be rewritten as the sum of pair-wise kernels:

\begin{equation}
    K_{iq} = \sum_{I \in i} \sum_{J \in q}  \;  K(\textbf{x}_I, \textbf{x}_{J})
    \label{eq:localkernel}
\end{equation}

with $I$ and $J$ being atoms in the training and query, respectively.
Kernels for local atomic representations of molecules are often augmented with an additional elemental screening function to compare only atoms of the same nuclear charge:

\begin{equation}
     K(\textbf{x}_I, \textbf{x}_J) = \delta_{Z_I, Z_J} \exp\left(-\frac{||\textbf{x}_I - \textbf{x}_J||^2_2}{2\sigma^2}\right)
    \label{eq:gaussfchl}
\end{equation}

with Kronecker Delta $\delta$, nuclear charges $Z$ and atomic representations \textbf{x}.
In order to use local representations as a normalized molecular similarity measure, Eq.\ref{eq:gaussfchl} can be expressed as the sum of local atomic comparisons:

\begin{equation}
     d_\textrm{global}(\textbf{x}_i, \textbf{x}_q) = \delta_{Z_I, Z_J} \sum_{I \in i} \sum_{J \in q} \textbf{x}_I \cdot \textbf{x}_J
    \label{eq:globaldistance}
\end{equation}

\begin{equation}
     K(\textbf{x}_i, \textbf{x}_q) = \exp\left(-\frac{d(\textbf{x}_i, \textbf{x}_i) - 2~d(\textbf{x}_i, \textbf{x}_q) + d(\textbf{x}_q, \textbf{x}_q))}{2\sigma^2}\right)
    \label{eq:gaussfchlglobal}
\end{equation}

\subsection{Representations}

We use standard representations which have been developed, benchmarked and continuously improved over the past years for learning quantum properties.
The Coulomb Matrix (CM\cite{rupp_fast_2012}) contains a 1-body self interaction and a 2-body Coulomb repulsion term and has been one of the first representations developed for machine learning quantum properties. 
The bag-of-bonds (BoB\cite{hansen_machine_2015}) representation, the direct successor of the CM, vectorizes the CM terms into fixed size bins, thus enabling the strict comparison between chemically similar environments.
The spectrum of London and Axilrod–Teller–Muto (SLATM\cite{AMONS}) representation uses London dispersion contributions as the 2-body and the Axilrod–Teller–Muto potential as the 3-body term and has shown improved performance over BoB for learning quantum properties.
The Faber–Christensen–Huang–Lilienfeld (FCHL19\cite{christensen_fchl_2020}) representation describes a local atomic environment per atom, respectively, and contains 2 and 3-body terms similar to Behler \& Parrinello atomic symmetry functions (ACSF\cite{behler_atom-centered_2011}) accounting for distance and angular information. 
We used the optimized FCHL19 parameters as recommended in Ref.\cite{christensen_fchl_2020}. 
The Smooth Overlap of Atomic Positions (SOAP)\cite{SOAP} representation is, similar to FCHL19, a local atomic representation that encodes local geometries through spherical harmonics and radial basis functions. We used $n_{max}=3$ and $l_{max}=5$ with $\sigma=0.1$ and a cutoff of 3.5 $\si{\angstrom}$ as SOAP settings.
For the learning of the free energy of solvation, we used FCHL19 and the free-energy-machine-learning (FML) scheme\cite{Weinreich2021}.

\subsection{Training}

We optimize the hyperparameters $\sigma$ and $\lambda$ through cross-validation on the largest training set size. 
Both parameters are held constant for all other training set sizes. 
To train an SML model, the pair-wise distances of a query to all available datapoints defined by $N_{max}$ are calculated.
An overview of training/query set sizes and distance metrics for each representation and dataset can be seen in Tab.~S1 in the SI.
After sorting all points by distance, only the $\overline{N}_q$ closest neighbours are considered for training.
SML learning curve are obtained by averaging across all single query runs.
Similarly, k-NN baselines are calculated by averaging the labels of the closest $\overline{N}_q$ neighbours weighted by the distance to the query, respectively.
RML models are trained choosing $N$ data points at random.
The regression coefficients are obtained through Eq.\ref{eq:train} and the query label predicted via Eq.\ref{eq:krr} (Fig.~1~B).
Note, RML models are only trained once to predict the labels of all queries, whereas SML and k-NN models are trained on-the-fly for each query, respectively.

\subsection{Data}

To assess SML, several quantum based datasets have been considered.
The QM7 dataset totals 7'165 organic molecules up to seven heavy atoms (CONS)\cite{rupp_fast_2012, blum_970_2009}. Atomization energies and structures were calculated using the Perdew-Burke-Ernzerhof hybrid functional (PBE0) level of quantum chemistry\cite{rupp_fast_2012, blum_970_2009}.
The QM9\cite{ramakrishnan_quantum_2014} dataset is a hallmark benchmark in quantum machine learning. 
Containing 133'885 small organic molecules up to nine heavy atoms (CONF), properties were calculated at the B3LYP/6-31G(2df,p) level of quantum chemistry\cite{ramakrishnan_quantum_2014}.
Additionally, a subset of 6'095 C$_7$O$_2$H$_{10}$ constitutional isomers has been used as a toy case. \par
The Enamine REAL database is the largest enumerated database of synthetically available compounds containing over 4.5 billion molecules\cite{enamine_real, enamine_realdb}.
We have chosen 1'000 random query compounds of the REAL library containing 6-21 heavy atoms. 
For training, we used the Enamine REAL building blocks, which encompass over 128'856 compounds that are used to synthesize the Enamine REAL dataset. 
We applied a SMILES\cite{weininger_smiles_1988} standardization protocol and filtered all compounds containing B, Si or Sn elements, yielding a total number of 124'440 compounds.
3D coordinates and conformers for all compounds were generated with ETKDG\cite{riniker_better_2015}. 
A maximum of 25 conformers were further relaxed using GFN2-xTB\cite{bannwarth_gfn2-xtbaccurate_2019} and corresponding Boltzmann weights obtained through energy rankings.
As a training label, free energies of solvation were computed using the Reaction-Mechanism-Generator\cite{RMG, RMG2}. \par

\section{Acknowledgement}
Icons used in Figures~\ref{fig:banana},\ref{fig:enamine} and S16. were created by Freepik, \texttt{photo3idea\_studio}, Victoruler, Maxim Basinski Premium and Monkik of flaticon.com.
This project has received funding from the European Research Council (ERC) under the European Union’s Horizon 2020 research and innovation programme (grant agreement No. 772834).
This research is part of the University of Toronto’s Acceleration Consortium, which receives funding from the Canada First Research Excellence Fund (CFREF).
Part of this research was performed while GFvR was visiting the Institute for Pure and Applied Mathematics (IPAM), which is supported by the National Science Foundation (Grant No. DMS-1925919).

\section{Author Contributions}
DL wrote new software used in the work, produced all figures, performed the literature search, compiled the references. 
DL and GvR acquired the data. 
DL, GvR and OAvL conceived and planned the project, analyzed and interpreted the results, and wrote the manuscript.

\section{Data and Code Availability}
The QM9 dataset is available at \url{https://dx.doi.org/10.6084/m9.figshare.c.978904.v5}.
The QM7 dataset is available at \url{http://quantum-machine.org/datasets/}.
The Enamine REAL database and building blocks can be accessed at \url{https://enamine.net/compound-collections/real-compounds/real-database}.
The code to produce SML learning curves is available at \url{https://doi.org/10.5281/zenodo.7870923}.

\section{Conflict of Interest}
The authors have no conflict of interest.

\section{References}
\bibliography{references.bib}{}
\bibliographystyle{ieeetr}

\end{document}

\onecolumngrid
\begin{center}
  \textbf{\large Improved decision making with similarity based machine learning: Applications in chemistry} \\[.5cm]
  \textbf{\large Supplementary Information} 
  \\[.5cm]
  Dominik Lemm$^{1,2}$, Guido Falk von Rudorff$^{1}$ and O. Anatole von Lilienfeld$^{3,4,5}$\\[.1cm]
  {
  \itshape ${}^1$ University of Vienna, Faculty of Physics,  Kolingasse 14-16, AT-1090 Vienna, Austria\\
  \itshape ${}^2$ University of Vienna, Vienna Doctoral School in Physics, Boltzmanngasse 5, AT-1090 Vienna, Austria\\
  \itshape ${}^3$Departments of Chemistry, Materials Science and Engineering, and Physics, University of Toronto, St. George Campus, Toronto, ON, Canada\\
  \itshape ${}^4$Vector Institute for Artificial Intelligence, Toronto, ON, M5S 1M1, Canada \\
  \itshape ${}^5$Machine Learning Group, Technische Universit\"at Berlin and Institute for the Foundations of Learning and Data, 10587 Berlin, Germany\\
  }
  ${}^*$Electronic address: anatole.vonlilienfeld@utoronto.ca\\
(Dated: \today)\\[1cm]
\end{center}

\newpage

\section*{Supplementary Text}

\subsection*{Local machine learning}
Local machine learning methods such as k-nearest neighbours are known since the 1950s and have been shown to be increasingly useful when the training data is not evenly distributed and locality effects dominate the model performance.
Formally, such weighted neighbourhood schemes can be expressed as:
\begin{equation}
    y_{pred} = \sum^{N}_{i} W(x_{q}, x_i)y_i
\label{eq:weight}
\end{equation}
with weighting function $W$, query representation $x_q$ and $x_i$ and $y_i$ being the representation and label of prospective training points, respectively. 
In case of k-nearest neighbour regression, the weighting function is often formalized through a query distance based decay, e.g. $\frac{1}{d}$, weighting the importance of the local query vicinity higher.
Similar to k-nearest neighbours, kernel based machine learning algorithms such as kernel ridge regression can also be seen as a form of weighted neighbourhood learning through the use of specific kernel function, e.g. Gaussian or Laplacian kernels. 
The local importance weighting is then achieved through optimizing the kernel width $\sigma$ in
$ \exp\left(-\frac{||\textbf{x}_i - \textbf{x}_q||^2_2}{2\sigma^2}\right)$.
Thus, kernel functions without locality weighting do not allow for a direct interpretation as a local learning scheme. 
Moreover, new predictions are obtained through calculated regression coefficients (see section IV.A) rather than directly averaging the labels $y$.
Besides kernel based methods, also bagged decision tree based models, e.g. random forests\cite{Lin02randomforests}, can be interpreted as a local learning algorithm.
Whereas all prior examples relied on weighting through explicit distance calculations, locality in decision tree's is achieved through recursively partitioning the feature space to group similar features and labels together. 
In the case of random forests, an ensemble of multiple decision tree's are bagged together following the assumption that the prediction of many weak learners will be superior over a single tree. 
A prediction for an unknown query is made by averaging the prediction of all tree's, thereby employing a similar weighting effect as described by Eq.~\ref{eq:weight}.

\subsection*{Representation Analysis on QM7/QM9}

We applied the SML scheme to learning atomization energies of C$_{7}$O$_{2}$H$_{10}$ constitutional isomers, QM7, as well as QM9 compounds, respectively. Results in Fig.\ref{fig:qm9full} suggest that SML's superior data efficiency is consistent throughout datasets and representations, respectively. The performance comparison between multiple representations in Fig.\ref{fig:qm9full}C indicate a difference in SML offset depending on the representation. Interestingly, a low offset does not necessarily indicate a higher model accuracy at larger training set sizes. For example, the BoB SML curve in Fig.\ref{fig:qm9full}C has a similar offset as FCHL19, however, converges much earlier compared to a better performing representation such as SLATM. Similar learning curve offsets of the global representation BoB\cite{hansen_machine_2015} and FCHL19 indicate that a combined use of a global representation for fast similarity searches and a local representation for regression might be advisable. In analogy to the learning curve and neighbour analysis performed for FCHL19 in Fig.3B-D, we analysed two compounds predicted with BoB SML models (Fig.\ref{fig:bobshells}). Results suggest that nearest neighbours obtained via BoB resemble more chemical similarities than neighbours obtained with FCHL19 (Fig.3D), with BoB even being able to identify a tautomere as the closest neighbour.

\subsection*{FCHL19 Kernel Analysis}
Due to the prevalent use of the Gaussian kernel with the FCHL19 representation, we compared the performance of Laplacian and Gaussian kernels. Fig.\ref{fig:sup_qm9_fchl} indicates better performance of a Gaussian type kernel at larger training set sizes.
Comparing the kernel principal component analysis (PCA) of a local Laplacian and Gaussian FCHL19 kernel displays a well separated pattern for the Gaussian type kernel (Fig.\ref{fig:pca_local}), which could be an explanation for the performance difference.
To calculate nearest neighbour distances and similarities with FCHL19, a global Laplacian or Gaussian type kernel is required as defined by Eq.10.
The histogram of kernel distance distributions indicates a much wider distribution of similarity values in a Laplacian kernel compared to the Gaussian kernel (Fig.\ref{fig:kernel_dist}), with most of the values being larger than 0.9. 
The kernel PCA of both kernels, however, shows no clear difference (Fig.\ref{fig:pca_global}). Both kernels show a buried PCA pattern for query compounds with high nearest neighbour density (Fig.\ref{fig:pca_global}A-B), whereas for a low nearest neighbour density compound the pattern is more separated (Fig.\ref{fig:pca_global}C-D).
Moreover, comparing the PCA patters of SML and RML kernels, the former shows more systematic trends, whereas the latter is more randomly distributed (Fig.\ref{fig:pca_random}).

\subsection*{Shared SML Models for Multiple Queries}
To determine the data efficiency of SML in consideration of multiple queries, we train shared neighbourhood SML models by combining the local nearest neighbours of multiple queries to compile the total training set (Fig.\ref{fig:para3d}).
Thus, the number of nearest neighbours per query $N_{neighs}$ becomes a hyperparameter. Moreover, we introduce an additional hyperparameter $N_{filter}$ that only considers nearest neighbours which occur at least 1, 2, 3 or 4 times in a finite radius of multiple query neighbourhoods.
We apply this scheme to train shared SML models tailored for up to 1'000 boutique compound queries of the Enamine REAL library (Fig.\ref{fig:para2d}). Results suggest that for all considered $N_{queries}$ and $N_{neighs}$ per query, a combination can be found that reaches 1~kcal/mol chemical accuracy with less than 2'000 training points (Fig.\ref{fig:enamine_ntrain}). Furthermore, optimal hyperparameter combinations are mostly found for $N_{filter}=1$, underlining the importance of neighbour proximity. For example, given two queries, a neighbour within a large radius of both queries could be found ($N_{filter}=2$) which might dissimilar to both queries and, therefore, be less impactful for learning than a direct neighbour ($N_{filter}=1$), respectively. Considering the optimal training set sizes for multiple queries (Fig.\ref{fig:enamine_ntrain}), the time and cost estimate depicted in Fig.5C can be derived.

\section*{Supplementary Tables}

\begin{table*}[ht]
\centering
\caption{Overview of maximum training set size $N_{max}$, number of queries $N_{Queries}$ and representation/similarity metric combination used to identify nearest neighbours for each dataset, respectively.}
\begin{tabular}{@{}lrrll@{}}
\toprule
\multicolumn{1}{l}{Dataset} & $N_{max}$ & $N_{Queries}$ & Representation         & Similarity Metric           \\ \midrule
QM7                          &    5'680         &          1'421     & Coulomb Matrix         & L1-Norm                   \\\midrule
QM9 C$_7$O$_2$H$_{10}$       &    5'000         &          1'095     & Coulomb Matrix         & L1-Norm                   \\ \midrule
QM9                          &    133'685         &       200        & Coulomb Matrix         & L1-Norm                   \\
                             &             &               & BoB                    & L1-Norm                   \\
                             &             &               & SLATM                  & L1-Norm                   \\
                             &             &               & FCHL19                 & Gaussian Kernel \\
Enamine REAL                 &  124'440           &    1'000           & Morgan Fingerprint                  & Tanimoto                  \\
\bottomrule
\end{tabular}
\label{tab:sizes}

\end{table*}

\newpage
\section*{Supplementary Figures}

\begin{figure*}[ht]
    \centering
    \includegraphics[width=\textwidth]{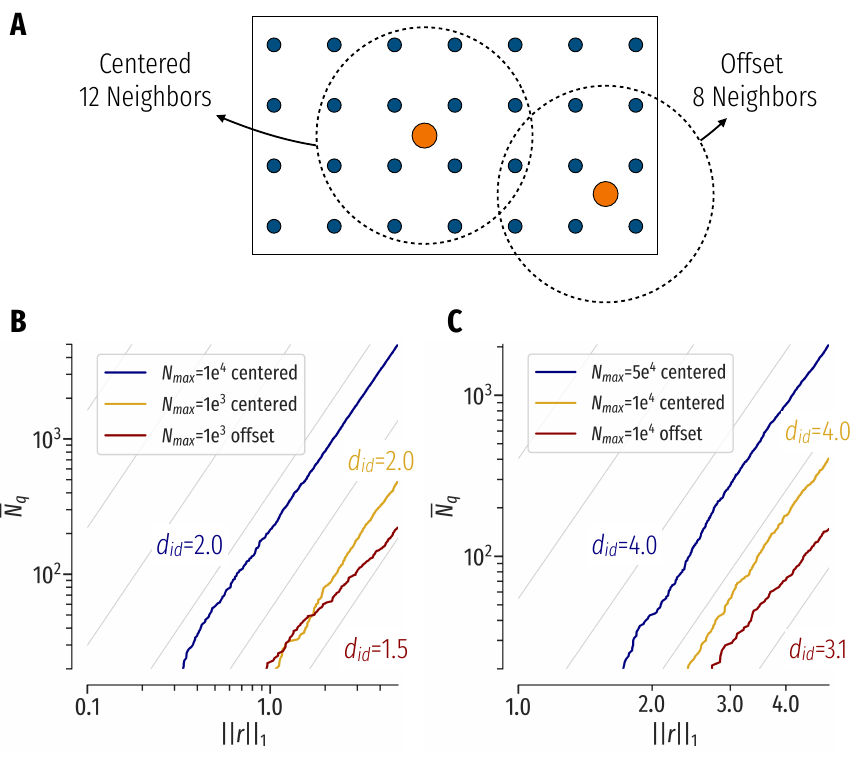}
    \caption{Toy example highlighting the effect of edge artifacts on intrinsic dimensionality ($d_{id}$) estimation.
    (\textbf{A}) Example of a 2D space with evenly spaced points (blue) and fixed boundaries. Given a fixed radius, a point centered in the middle will have more nearest neighbours than a point offset to the boundary.
    The latter leads to an underestimation of the $d_{id}$.
    (\textbf{B}) $d_{id}$ of centered and offset points in a fixed boundary 2D space.
    (\textbf{C}) $d_{id}$ of centered and offset points in a fixed boundary 4D space.
    }
    \label{fig:toy_offset}
\end{figure*}

\begin{figure*}[ht]
    \centering
    \includegraphics[width=\textwidth]{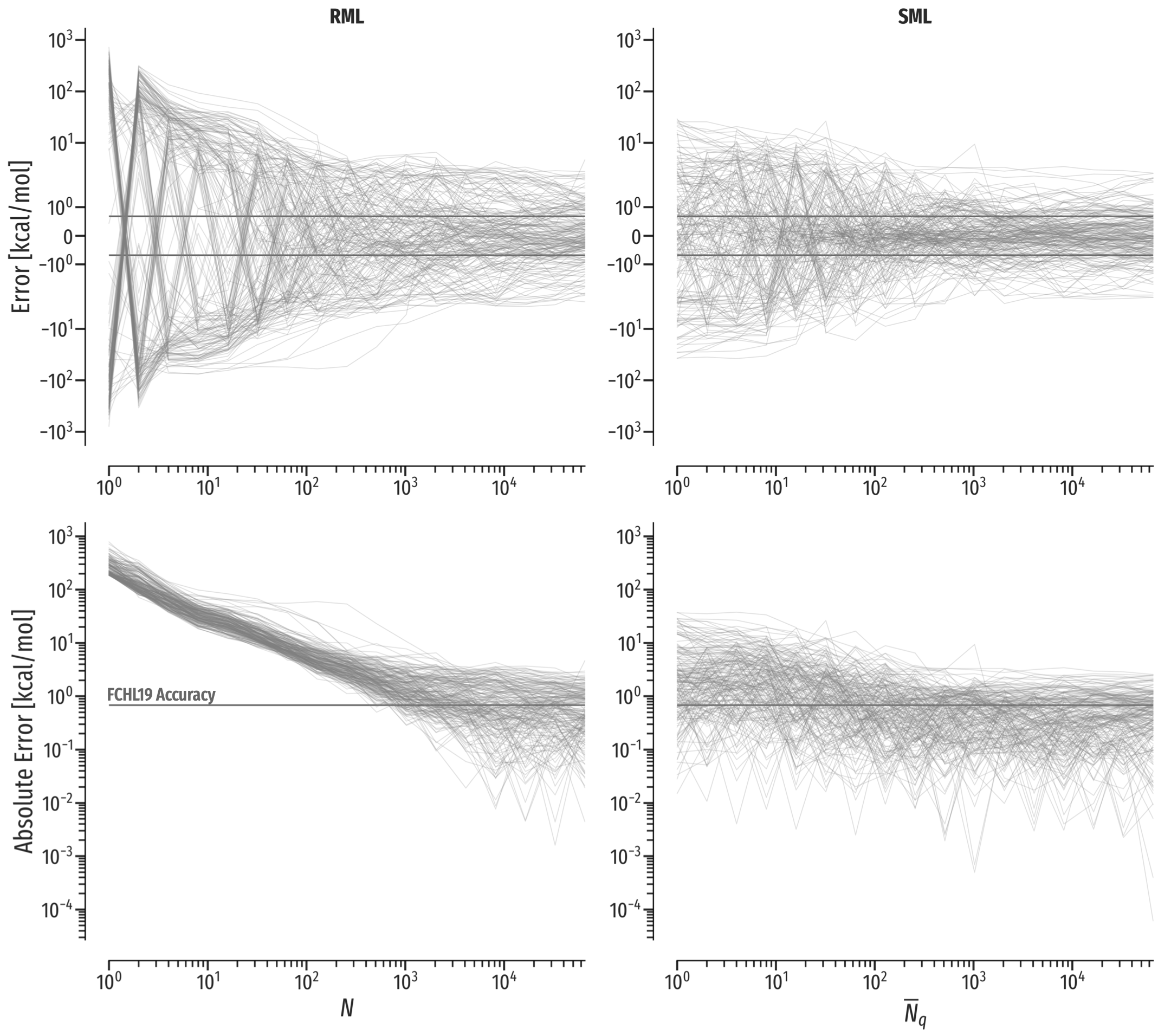}
    \caption{Randomly chosen atomization energy learning curves of 100 QM9\cite{ramakrishnan_quantum_2014} queries predicted via RML and SML using FCHL19\cite{christensen_fchl_2020}, respectively. The FCHL19 accuracy represents the MAE of an RML model at maximum training set size.}
    \label{fig:fchlsingle}
\end{figure*}

\begin{figure*}[ht]
    \centering
    \includegraphics[width=0.5\textwidth]{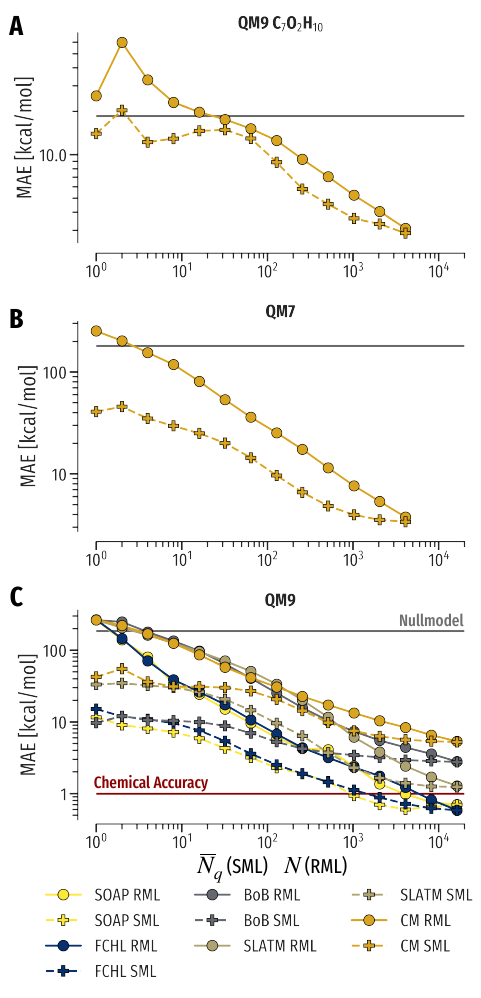}
    \caption{Learning curves of SML applied to learning atomization energies of C$_{7}$O$_{2}$H$_{10}$ constitutional isomers, QM7 and the QM9\cite{ramakrishnan_quantum_2014} dataset using different molecular representations, respectively.
    (\textbf{A}) Comparison between SML and RML learning curves applied to C$_{7}$O$_{2}$H$_{10}$ isomers using the Coulomb Matrix\cite{rupp_fast_2012} (CM) representation.
    (\textbf{B}) Comparison between SML and RML learning curves applied to the QM7 dataset using the Coulomb Matrix\cite{rupp_fast_2012} (CM) representation.
    (\textbf{C}) Comparison between SML and RML learning curves applied to QM9\cite{ramakrishnan_quantum_2014} using various representations.}
    \label{fig:qm9full}
\end{figure*}

\begin{figure*}[ht]
    \centering
    \includegraphics[width=0.5\textwidth]{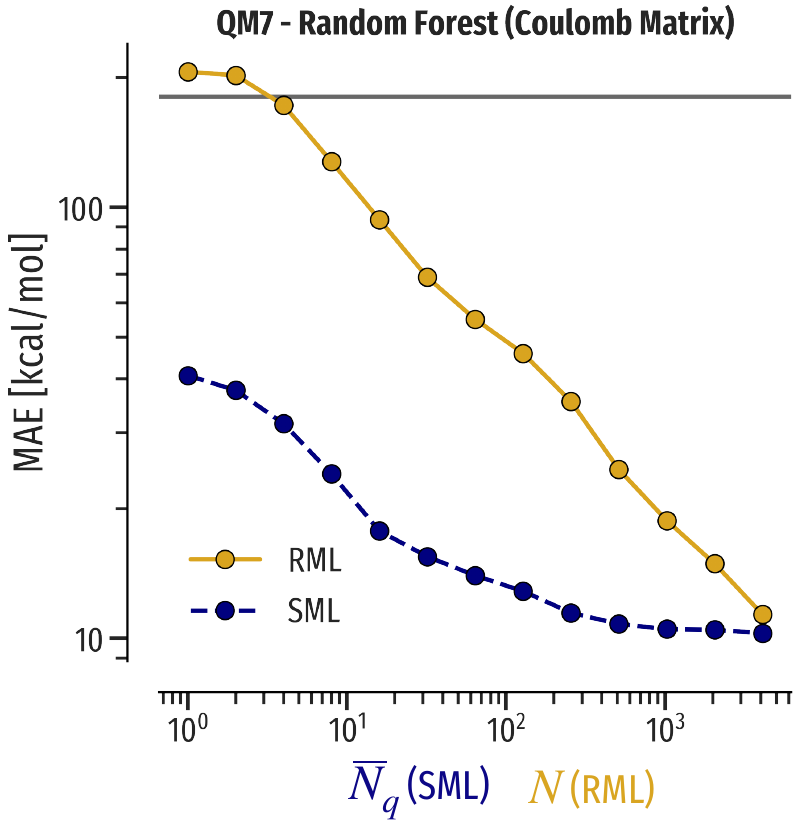}
    \caption{Learning curve of SML applied to learning atomization energies of QM7 compounds using the Coulomb Matrix\cite{rupp_fast_2012} molecular representations with a random forest regressor.}
    \label{fig:qm7rf}
\end{figure*}

\begin{figure*}[ht]
    \centering
    \includegraphics[width=\textwidth]{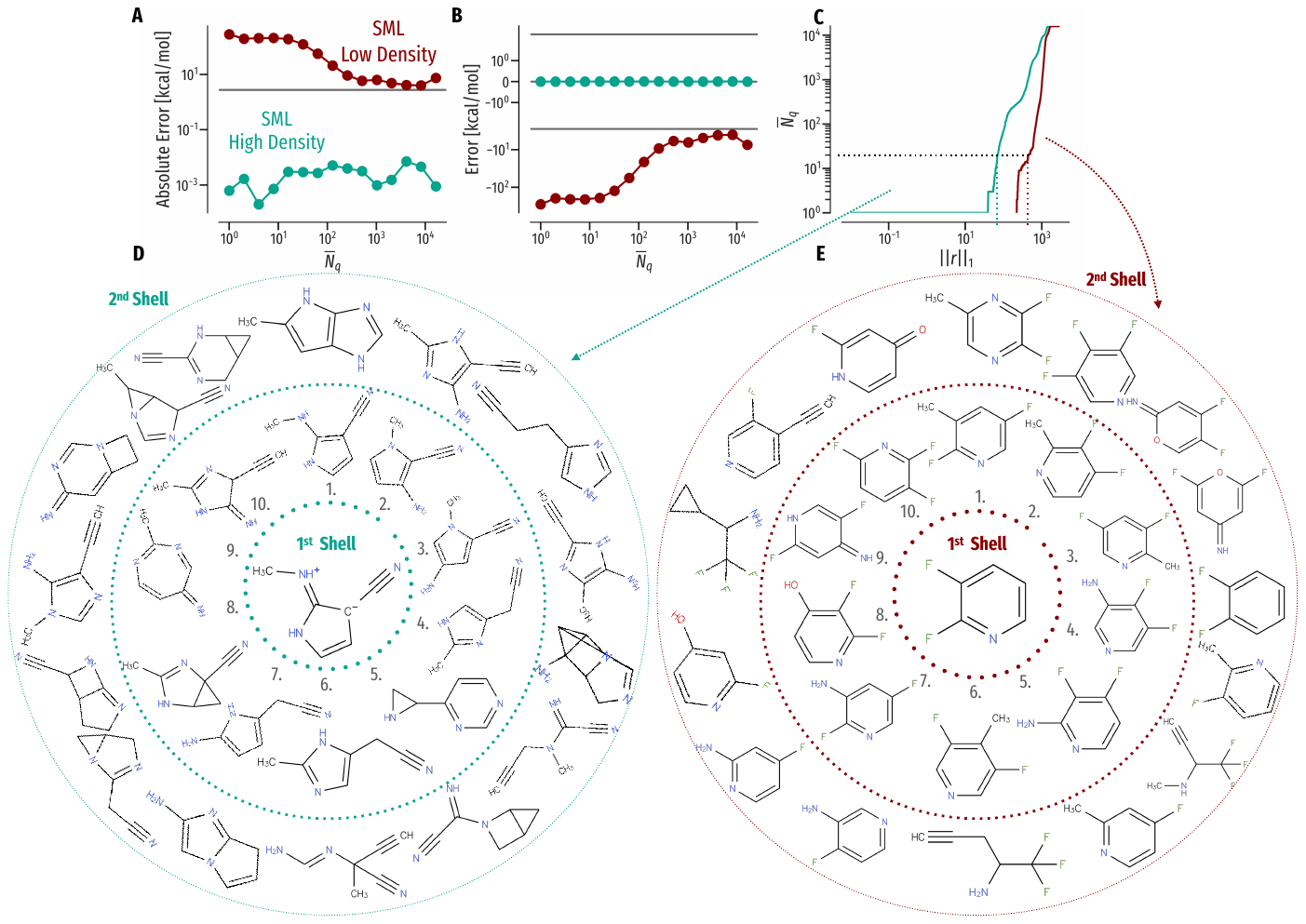}
    \caption{
    Analysis of the influence on neighbourhood density on learning atomization energies of two QM9 compounds with either low or high nearest neighbour density using the BoB representation, respectively.
    (\textbf{A-B}) Learning curves of SML models for predicting two compounds with either low or high nearest neighbour density, respectively. 
    (\textbf{C}) Amount of nearest neighbours found within an increasing distance radius, respectively.
    (\textbf{D-E}) Display of the first shell (closest ten molecules) and second shell (next closest fifteen) neighbours for the two query compounds.
}
    \label{fig:bobshells}
\end{figure*}

\begin{figure*}[ht]
    \centering
    \includegraphics[width=\textwidth]{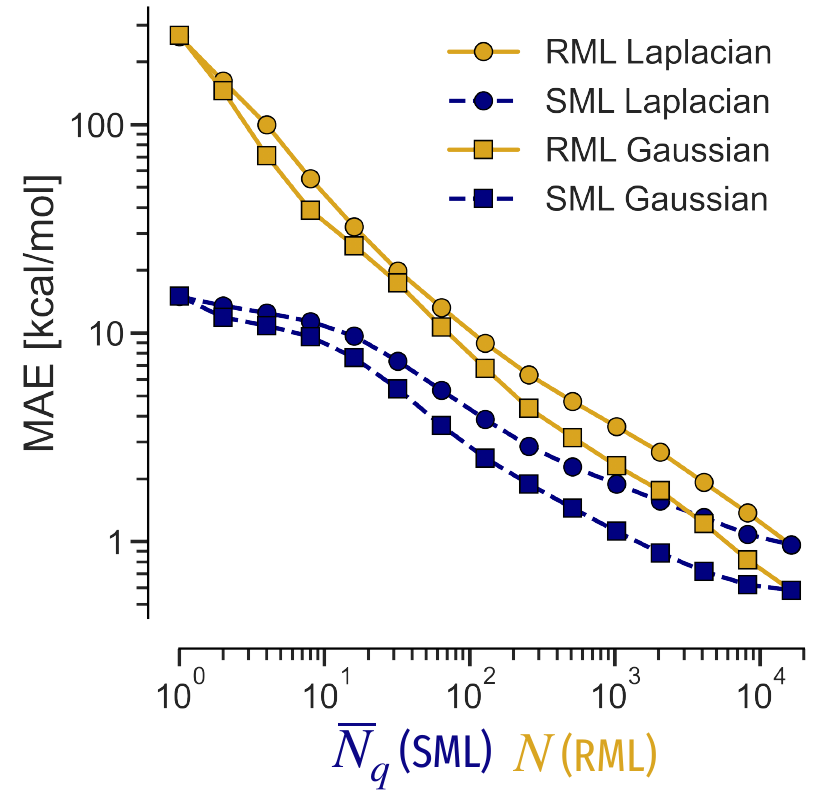}
    \caption{Performance comparison between a Laplacian and Gaussian FCHL19\cite{christensen_fchl_2020} kernel in combination with SML or RML to learn atomization energies of QM9\cite{ramakrishnan_quantum_2014} compounds, respectively.}
    \label{fig:sup_qm9_fchl}
\end{figure*}

\begin{figure*}[ht]
    \centering
    \includegraphics[width=\textwidth]{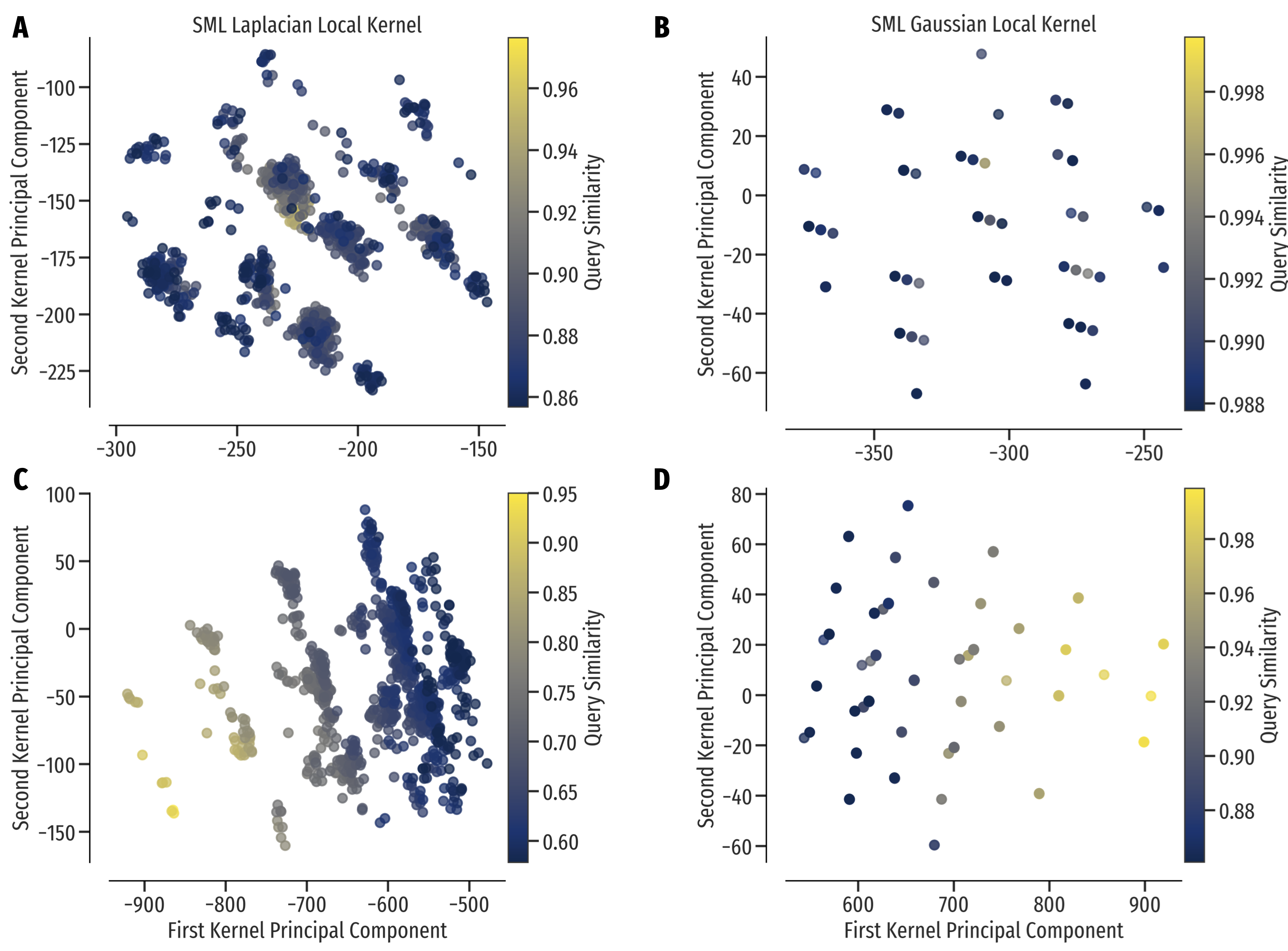}
    \caption{
    Comparison between kernel PCA's of local Laplacian and Gaussian SML kernels using the FCHL19\cite{christensen_fchl_2020} representation applied to a subset of QM9\cite{ramakrishnan_quantum_2014}, respectively.
    (\textbf{A}) SML Laplacian kernel PCA of a compound with high neighbour density.
    (\textbf{B}) SML Gaussian kernel PCA of a compound with high neighbour density.
    (\textbf{C}) SML Laplacian kernel PCA of a compound with low neighbour density.
    (\textbf{D}) SML Gaussian kernel PCA of a compound with low neighbour density.
    }
    \label{fig:pca_local}
\end{figure*}

\begin{figure*}[ht]
    \centering
    \includegraphics[width=\textwidth]{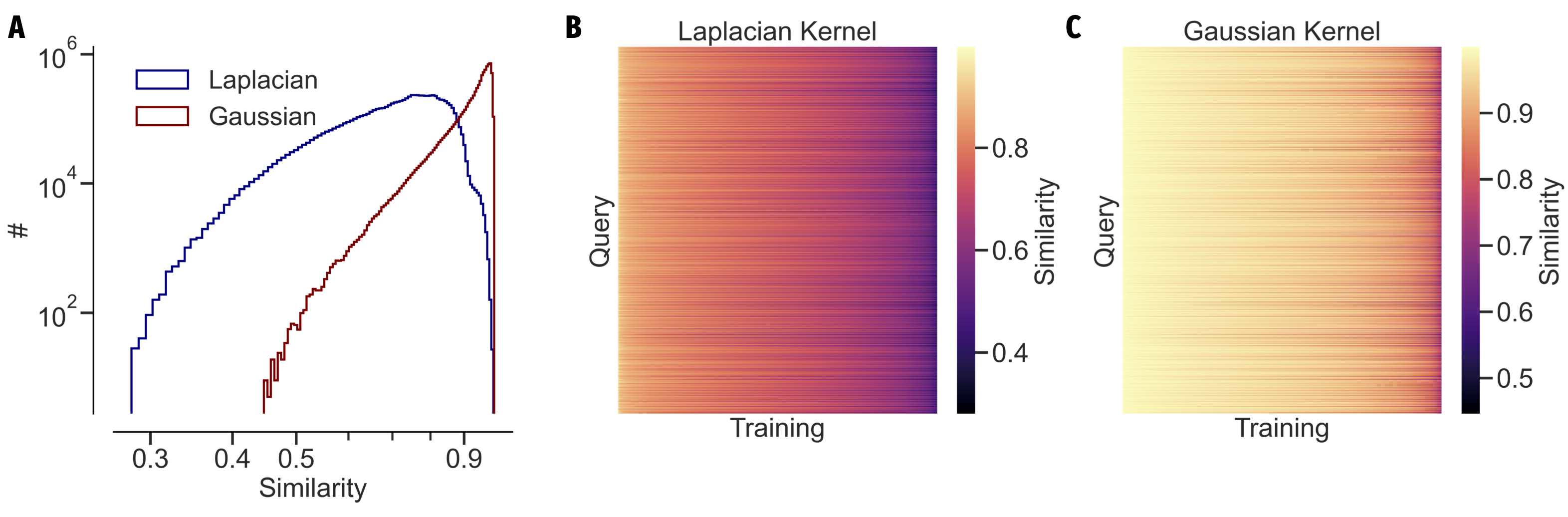}
    \caption{
    Comparison between the distance/similarity distribution of a global  Gaussian or Laplacian kernel using the FCHL19\cite{christensen_fchl_2020} representation on a subset of QM9\cite{ramakrishnan_quantum_2014}, respectively.
    (\textbf{A}) Histogram of kernel similarities of a global Gaussian and Laplacian kernel.
    (\textbf{B}) Heatmap of sorted kernel similarities per query using a Laplacian kernel.
    (\textbf{C}) Heatmap of sorted kernel similarities per query using a Gaussian kernel.
    }
    \label{fig:kernel_dist}
\end{figure*}

\begin{figure*}[ht]
    \centering
    \includegraphics[width=\textwidth]{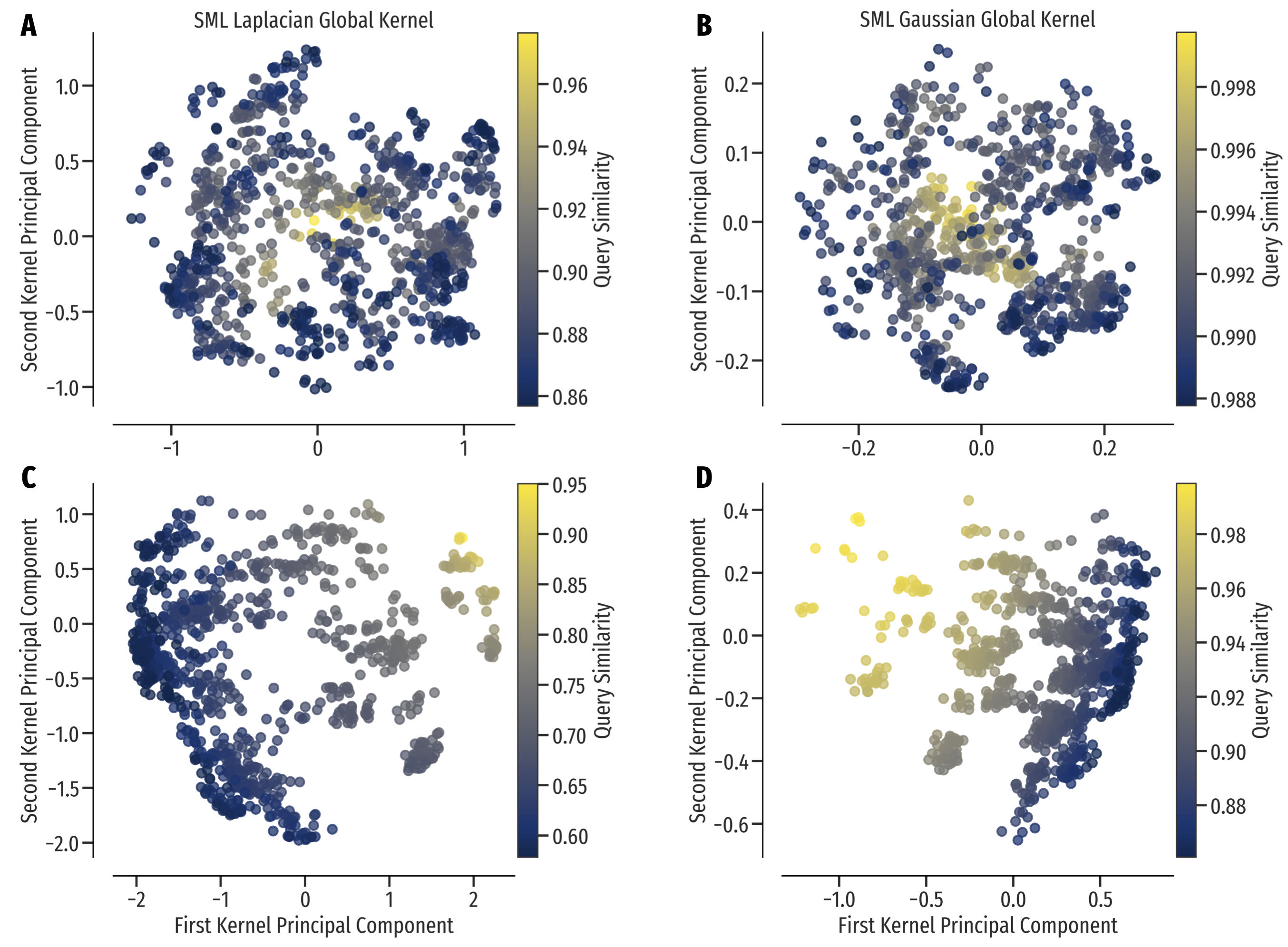}
    \caption{Comparison between kernel PCA's of global Laplacian and Gaussian SML kernels using the FCHL19\cite{christensen_fchl_2020} representation applied to a subset of QM9\cite{ramakrishnan_quantum_2014}, respectively.
    (\textbf{A}) SML Laplacian kernel PCA of a compound with high neighbour density.
    (\textbf{B}) SML Gaussian kernel PCA of a compound with high neighbour density.
    (\textbf{C}) SML Laplacian kernel PCA of a compound with low neighbour density.
    (\textbf{D}) SML Gaussian kernel PCA of a compound with low neighbour density.
    }
    \label{fig:pca_global}
\end{figure*}

\begin{figure*}[ht]
    \centering
    \includegraphics[width=\textwidth]{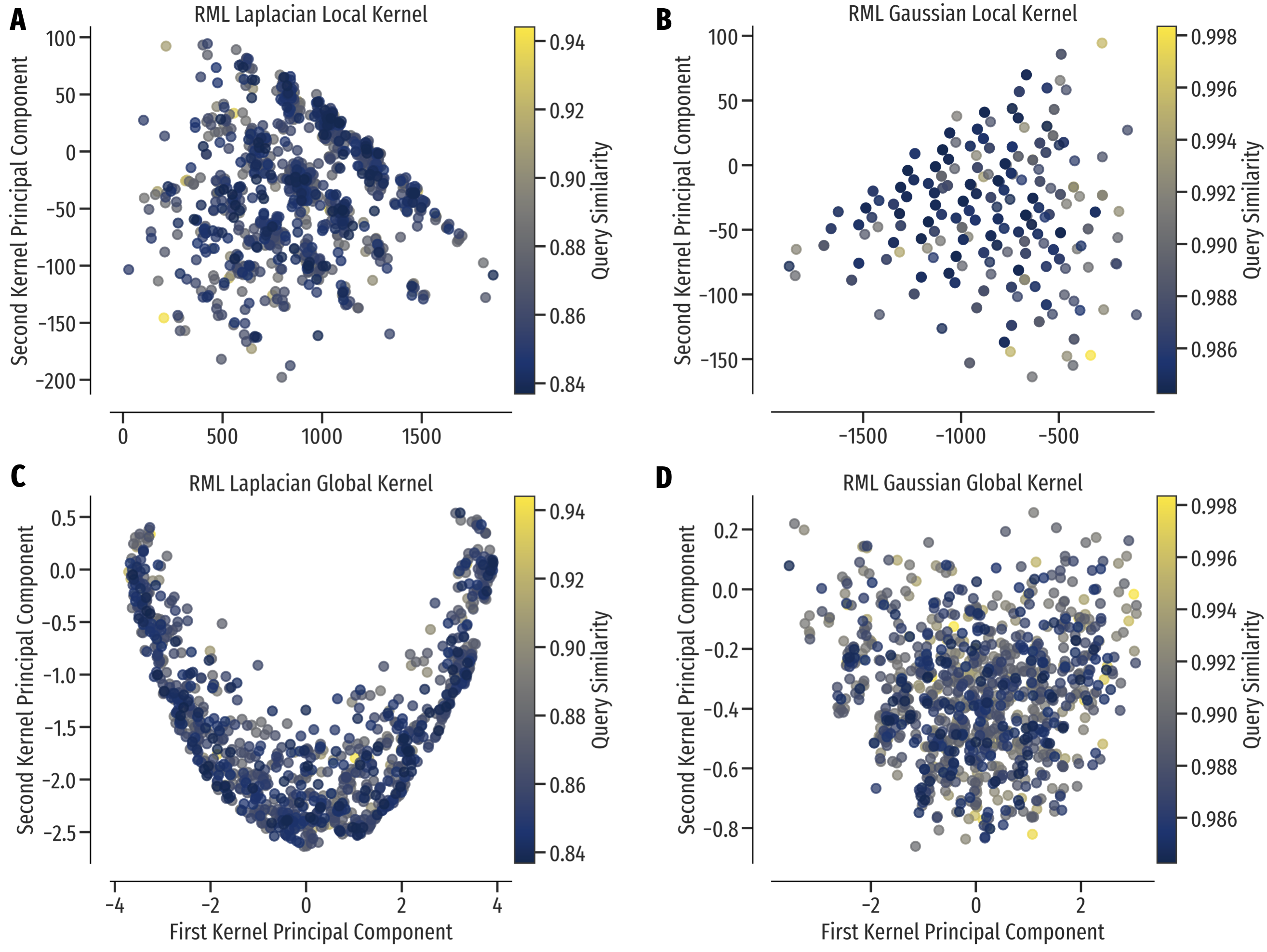}
    \caption{Comparison between kernel PCA's of global Laplacian and Gaussian RML kernels using the FCHL19\cite{christensen_fchl_2020} representation applied to a subset of QM9\cite{ramakrishnan_quantum_2014}, respectively.
    (\textbf{A}) Local Laplacian kernel PCA.
    (\textbf{B}) Local Gaussian kernel PCA.
    (\textbf{C}) Global Laplacian kernel PCA.
    (\textbf{D}) Global Gaussian kernel PCA.
    }
    \label{fig:pca_random}
\end{figure*}

\begin{figure*}[ht]
    \centering
    \includegraphics[width=\textwidth]{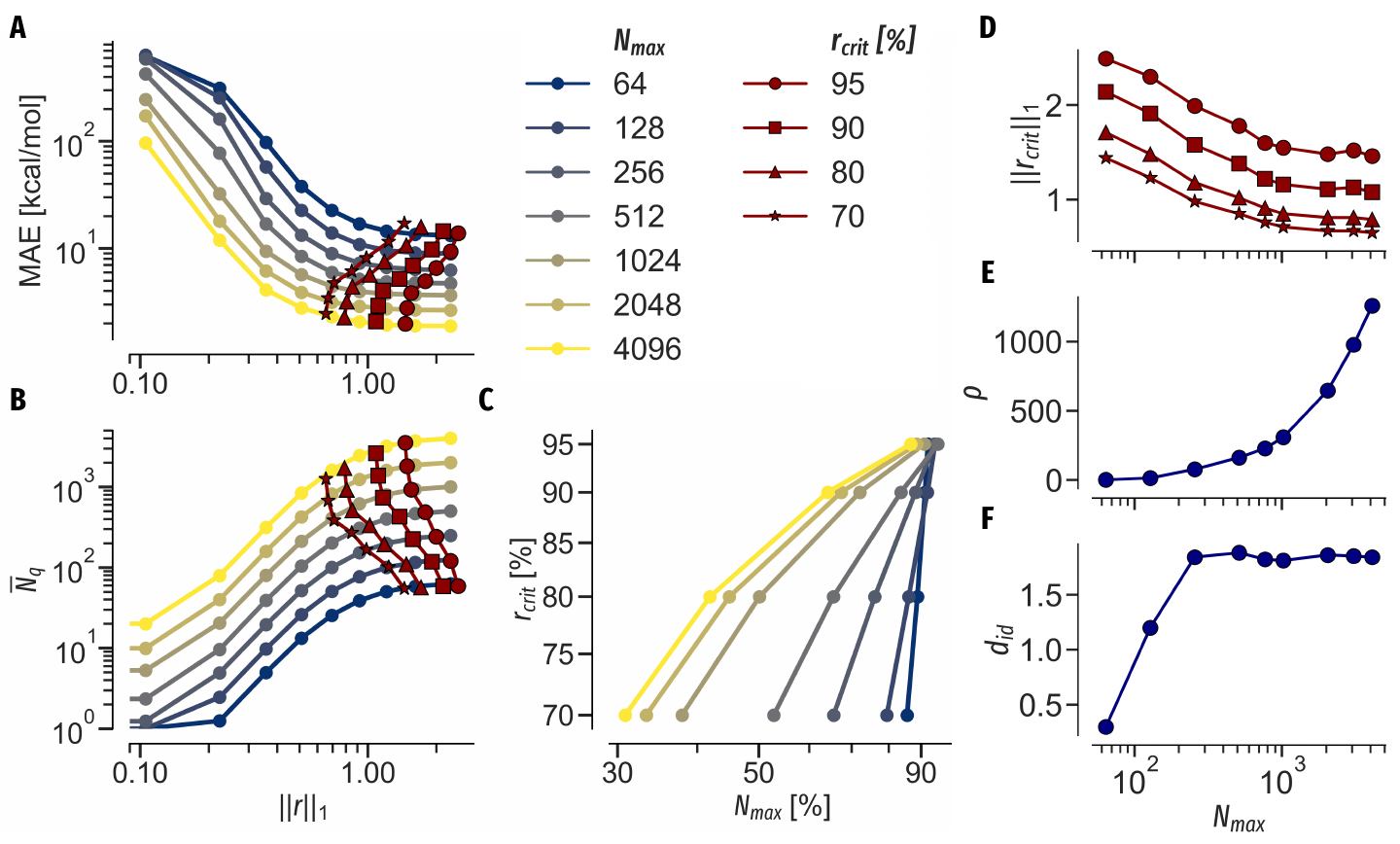}
    \caption{Distance based analysis of SML models of learning atomization energies of QM9 using FCHL19 and a Laplacian kernel. 
    $r_{crit}$ represents the distance radius at which a certain percentage of the maximally attainable predictive accuracy is reached.
    (\textbf{A}) Systematic learning of atomization energies using only training instances within the L1-Distance specified on abscissa.  
    (\textbf{B}) Averaged number of nearest neighbours around a query at different total $N_{max}$ for increasing L1-distance radius. 
    (\textbf{C}) Percentage of data set size necessary to reach a certain accuracy as determined via $r_{crit}$.
    (\textbf{D-F}) $r_{crit}$, density and intrinsic dimensionality ($d_{id}$) of the QM9\cite{ramakrishnan_quantum_2014} chemical space as a function of available training points, respectively. 
    }
    \label{fig:laplacian_sml}
\end{figure*}

\begin{figure*}[ht]
    \centering
    \includegraphics[width=\textwidth]{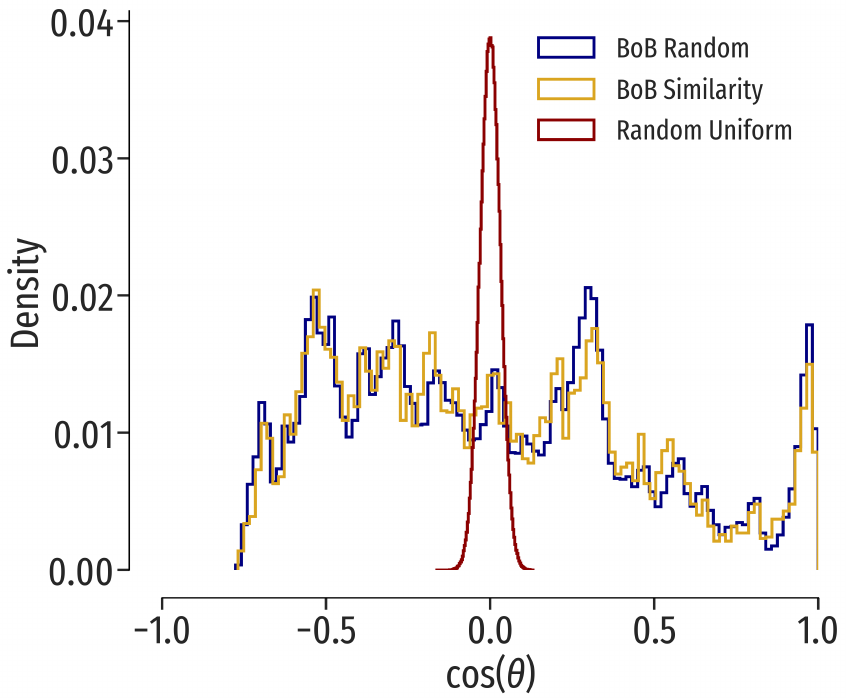}
    \caption{Histogram of calculated angles between 10'000 QM9\cite{ramakrishnan_quantum_2014} BoB\cite{hansen_machine_2015} representation vectors (approximate vector length $\sim$1'000) chosen at random or by similarity, respectively. 
    As a comparison, angles between 10'000 random uniform vectors (vector length 1'000) are shown in red.}
    \label{fig:bob_angles}
\end{figure*}

\begin{figure*}[ht]
    \centering
    \includegraphics[width=\textwidth]{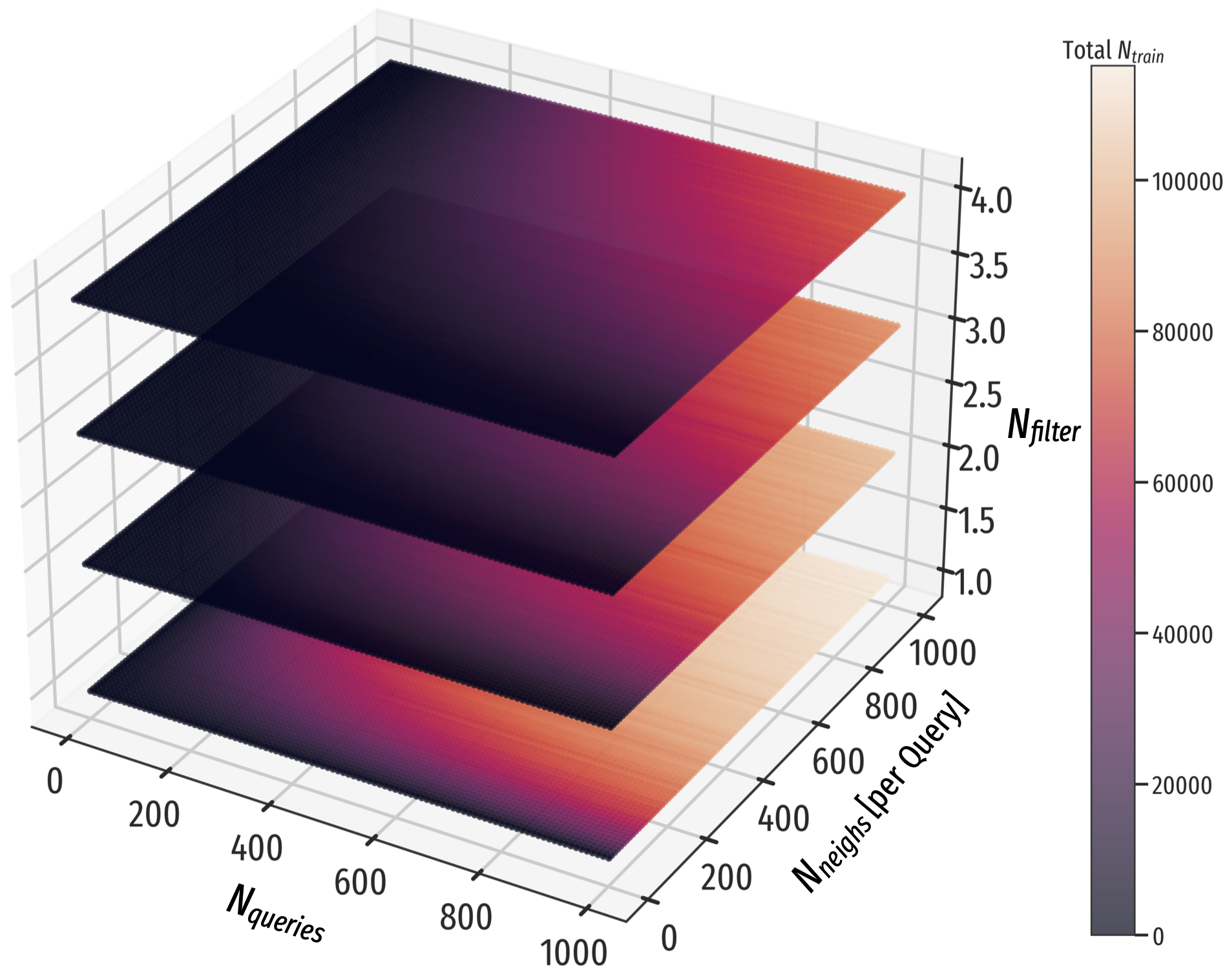}
    \caption{Visualization of the total number of training points $N_{train}$ as a result of sharing neighbourhoods between queries in the Enamine REAL dataset.
    To construct a shared SML model for an anticipated number of queries $N_{queries}$, a certain number of nearest neighbours $N_{neighs}$ per query have to be chosen to compile a training set. To reduce $N_{train}$ and $N_{neighs}$, an additional filter ($N_{filter}$) can be considered to only include neighbours that occur at least 1, 2, 3 or 4 times in multiple query neighbourhoods.}
    \label{fig:para3d}
\end{figure*}

\begin{figure*}[ht]
    \centering
    \includegraphics[width=\textwidth]{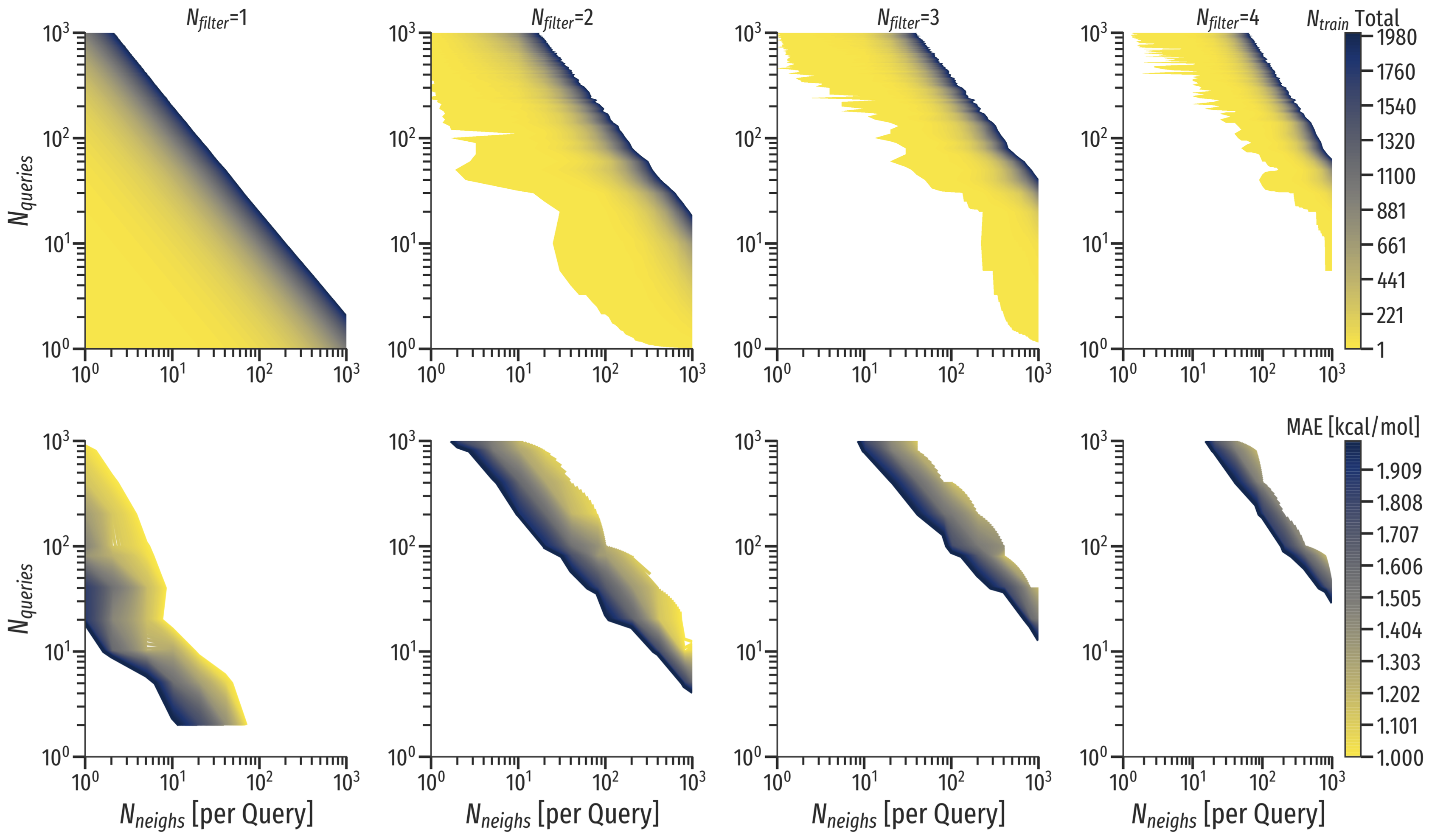}
    \caption{
    Analysis of the optimal combination of $N_{neighs}$ and $N_{filter}$ per query in order to reach chemical accuracy (1~kcal/mol) in learning free energies of solvation of queries in the Enamine REAL dataset via SML. $N_{filter}$ only includes neighbours that occur at least 1, 2, 3 or 4 times in multiple query neighbourhoods.
    The upper row depicts the boundary for number of training points when an SML would be more efficient than RML. The latter reaches chemical accuracy at $N_{train}$=2'000 (see Fig.4B).
    The lower row depicts shared SML model accuracy resulting from a grid search through various $N_{queries}$, $N_{neighs}$ and $N_{filter}$ combinations below the boundary of $N_{train}$=2'000.}
    \label{fig:para2d}
\end{figure*}

\begin{figure*}[ht]
    \centering
    \includegraphics[width=\textwidth]{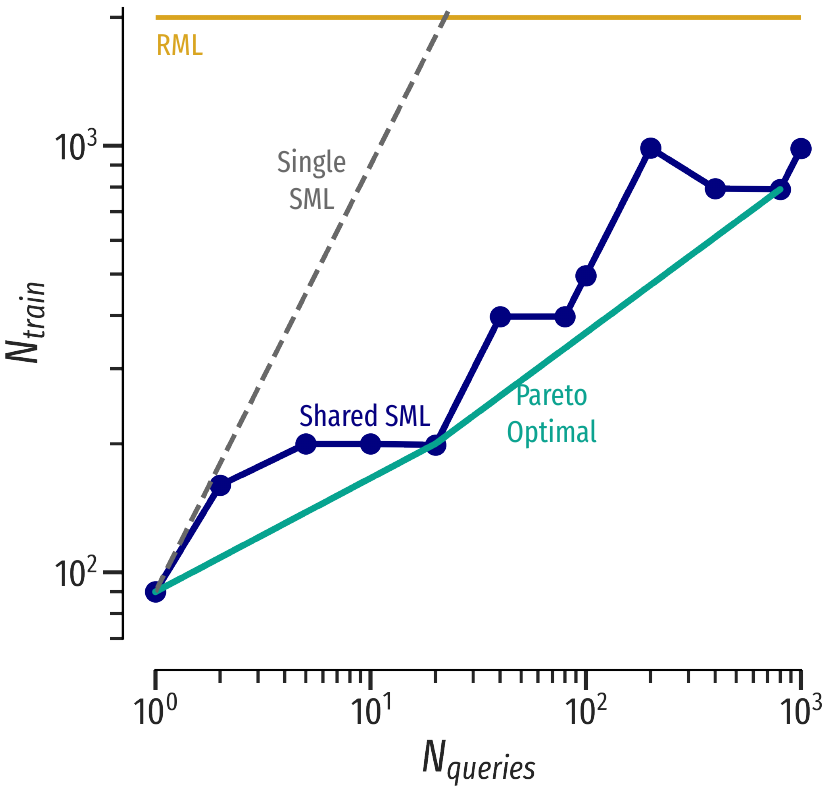}
    \caption{Total number of training points needed to generate free energy of solvation SML/RML models that reach chemical accuracy for Enamine REAL compounds as a function of number of queries.}
    \label{fig:enamine_ntrain}
\end{figure*}

\clearpage

\section*{Supplementary References}

\bibliography{references.bib}{}
\bibliographystyle{ieeetr}